\def\BE{\begin{equation}}
\def\EE#1{\label{#1}\end{equation}}
\def\U{{\cal U}}
\def\p{{\mathfrak P}}
\def\D{\,{\rm d}}
\def\e{{\rm e}}
\def\i{{\rm i}}
\def\bxi{{\bm\xi}}
\def\rf#1{(\ref{#1})}
\def\LA{\mbox{$\langle\hspace*{-.25em}\langle$}}
\def\RA{\mbox{$\rangle\hspace*{-.25em}\rangle$}}
\def\LLA{\mbox{$\Big\langle\hspace*{-.4em}\Big\langle$}}
\def\RRA{\mbox{$\Big\rangle\hspace*{-.4em}\Big\rangle$}}
\def\la{\langle}
\def\ra{\rangle}
\def\lb{\left\{}
\def\rb{\right\}}
\documentclass[12pt]{Melsarticle}
\usepackage{amsmath,amssymb,textcomp,bm,graphicx,mathrsfs}
\begin{document}
\renewcommand{\l}{{\mathfrak L}}
\title{Large-scale weakly nonlinear perturbations
of convective magnetic dynamos in a rotating layer}
\author[ita]{R.~Chertovskih}
\author[mitpan]{V.~Zheligovsky}
\address[ita]{Institute of Aeronautical Technology,\\
Pra\c{c}a Marechal Eduardo Gomes 50, Vila das Ac\'acias,
12228-900 S\~ao Jos\'e dos Campos, SP -- Brazil}
\address[mitpan]{Institute of Earthquake Prediction Theory and
Mathematical Geophysics, Russian Ac. Sci.,\\
84/32 Profsoyuznaya St, 117997 Moscow, Russian Federation}

\begin{abstract}
We present a new mechanism for generation of large-scale magnetic field
by thermal convection which does not involve the $\alpha$-effect.
We consider weakly nonlinear perturbations of space-periodic steady convective
magnetic dynamos in a rotating layer that were identified in our previous work.
The perturbations have a spatial scale in the horizontal direction that is much
larger than the period of the perturbed convective magnetohydrodynamic
state. Following the formalism of the multiscale stability theory, we have
derived the system of amplitude equations governing the evolution of
the leading terms in the expansion of the perturbations
in power series in the scale ratio. This asymptotic analysis is more involved
than in the cases considered earlier, because the kernel of the operator
of linearisation has zero-mean neutral modes whose origin lies in the spatial
invariance of the perturbed regime, the operator reduced on the generalised
kernel has two Jordan normal form blocks of size two, and simplifying symmetries
of the perturbed state are now missing. Numerical results for the amplitude
equations show that a large-scale perturbation, periodic in slow horizontal
variable, either converges to a short-scale neutral stability mode with
amplitudes tending to constant values, or it blows up at a finite slow time.
\end{abstract}

\maketitle
\section{Introduction}\label{intro}

A fundamental problem in astrophysics is to understand the sources of magnetic
fields that are featured by many astrophysical bodies such as the Sun and
the Earth. It is generally accepted that the magnetic fields are generated
by hydromagnetic processes in the melted or fluid-like interiors of the bodies.
This idea goes back to J.~Larmor \cite{La1,La2}. It is widely believed
that the so-called $\alpha$-effect plays a prominent role in such processes.
Such a mechanism of magnetic field generation was first suggested by E.~Parker
\cite{Pa57,Pa83}. It relies on the frozenness of magnetic field into
the conducting medium, when magnetic diffusion is negligible, implying that
a small eddy in turbulent flow deforms the magnetic field force line
into a loop. If this effect does not disappear when averaged over many loops,
it gives rise to a mean electromotive force (e.m.f.), which can be parallel
to the mean unperturbed magnetic field and can amplify the original mean field.

The theory of mean-field electrodynamics (MFE) is developed around a similar
central idea \cite{St,Kr}: Suppose the flow and magnetic field are split into
the mean and fluctuating parts, and the interaction of the fluctuating parts
of the flow and the field yields a non-zero mean e.m.f. The MFE
theory postulates that, at least in the case of magnetohydrodynamic (MHD)
turbulence, the latter is related to the mean magnetic field
(the coefficient of proportionality is traditionally denoted by $\alpha$,
and, accordingly, the phenomenon is called the ``$\alpha$-effect'') combined,
in some MFE models, with the spatial gradient of the mean field.

The MFE theory does not fully justify such relations. An insight into
their mathematical roots is provided by the multiscale stability theory (MST).
It treats an idealised case, in which turbulence is modelled by a small-scale
laminar flow, periodic in space and time (see \cite{vz}). In contrast with MFE,
mathematically rigorous results are then obtained by applying the asymptotic
theory of PDE homogenisation to the problem of linear or weakly nonlinear
stability of small-scale states. Purely hydrodynamic \cite{DF,GVF}
perturbations, the kinematic dynamo problem \cite{La,ZPF,ZP} (which is
an instance of the general linear MHD stability problem,
where the perturbed state is amagnetic, the flow and magnetic components
of the perturbation therefore decouple, and one focuses on the magnetic
perturbation), and full perturbations of forced MHD or
convective MHD states \cite{Ba,Z06,Z07,Z08} (see also \cite{vz}, Chap.~6--9)
were considered. Perturbations are supposed to involve spatial and temporal
scales that are much larger than the respective periods of the perturbed states.
The perturbations are linear combinations of amplitude-modulated small-scale
modes (i.e., eigenfunctions of the linearisation whose periods coincide with
those of the perturbed states) associated with the same eigenvalue. The coefficients
depend exclusively on slow temporal and spatial variables. They are usually
called amplitudes, and the equations governing their dynamics are called
amplitude equations. The evolution of such large-scale perturbations is essentially
controlled by the associated eigenvalue of the constituting small-scale modes
--- only neutral (belonging to the kernel of linearisation) small-scale modes
can instigate instability in the presence of large scales; consequently,
MST usually focuses on large-scale amplitude modulation of neutral modes. Since
typically small-scale neutral modes have non-zero means, amplitude equations
for large-scale perturbations of forced MHD states (convective or not) are
mean-field equations similar to those considered in MFE theory. However, as we
will see, this is not always the case: translation-invariant physical systems,
such as free thermal hydromagnetic convection in a horizontal layer investigated
in the present paper, possess zero-mean neutral small-scale stability modes.
Consequently, the mean-field description of the dynamics of large-scale
perturbations of such physical systems is inadequate.

We intend to carry out a detailed investigation of stability to large-scale
weakly nonlinear perturbations of the steady and time-periodic regimes
of magnetic field generation by free thermal convection of rotating electrically
conducting fluid in a horizontal layer, that were determined numerically
in \cite{CGPZ} and \cite{Z10}. Here we consider space-periodic convective MHD
steady states constituting the branch S$^{\rm R1}_8$ \cite{CGPZ}.
The group of symmetries of this steady state is generated by the symmetry
about the vertical axis, $x_3$, (note that this is not axisymmetry,
see section \ref{equ}) \cite{CGPZ,vz} and
the superposition of reflection about the midplane with translation by half
a period in the horizontal direction $x_1$. This group of symmetries is
smaller than other ones, typical for the parameter values considered
{\it ibid.}, and the system of amplitude equations, derived in \cite{vz},
is inapplicable for large-scale perturbations of states having this
group. The symmetry about a vertical axis implies that
the steady state does not possess the $\alpha$-effect. The large-scale
dynamics is due to the interplay of the combined eddy diffusivity and eddy
advection. We find that their interaction cannot sustain stationary
generation of large-scale magnetic field: a large-scale perturbation
from the class which is described by the multiscale formalism either converges
to a short-scale neutral stability mode, or it blows up at a finite time.

The paper is organised as follows. In section~\ref{equ} we derive the system
of amplitude equations for steady states of free hydromagnetic convection,
which have the same group of symmetries as those comprising the branch
S$^{\rm R1}_8$. In section~\ref{resu} we present results of numerical analysis
of the system of amplitude equations for several states belonging to this
branch. Finally, we make remarks triggered by our investigation.

\section{The multiscale formalism for large-scale perturbations\\
of free hydromagnetic convection}\label{equ}

In this section we derive amplitude equations for large-scale
perturbations of small-scale steady free hydromagnetic convection
in a horizontal layer of electrically conducting fluid rotating
about the vertical axis. A field, depending only on the fast spatial variables,
$\bf x$, and time, $t$, is called small-scale; if, in addition, the field
depends on the slow horizontal spatial variables, ${\bf X}=\varepsilon(x_1,x_2)$
and on the slow time, $T=\varepsilon^st$, where $s\ge 1$, it is called
large-scale. We will use asymptotic methods that are standard within the MST
approach. We assume that the perturbed state is symmetric about the vertical
axis and has equal periods $\mathscr P$ in $x_1$ and $x_2$; further assumptions
are introduced where they become relevant.

A three-dimensional field $\bf f$ is called symmetric about a vertical axis
passing through point $(a_1,a_2,0)$ when the following conditions are satisfied
\cite{CGPZ,vz}:
\begin{align*}
f_1(a_1-x_1,a_2-x_2,x_3)&=-f_1(a_1+x_1,a_2+x_2,x_3),\\
f_2(a_1-x_1,a_2-x_2,x_3)&=-f_2(a_1+x_1,a_2+x_2,x_3),\\
f_3(a_1-x_1,a_2-x_2,x_3)&=f_3(a_1+x_1,a_2+x_2,x_3).
\end{align*}
In particular, this symmetry implies that the flow is vertical everywhere
on the axis. Together with the $\mathscr P$-periodicity in the horizontal
directions, the symmetry about the Cartesian axis $x_3$ implies that the field
is also symmetric about vertical axes through points ${\mathscr P}{\bf n}/2$,
where ${\bf n}=(n_1,n_2,0)$ has integer components.

\subsection{Small-scale convective hydromagnetic steady states
and their perturbations}\label{state}

The state, whose stability we examine, is governed by the Navier--Stokes,
magnetic induction and heat transfer equations. In the coordinate system,
co-rotating with the fluid layer about the axis $x_3$, they are \cite{Cha}
\begin{align}
\partial{\bf V}/\partial t=\,&
\nu\nabla^2{\bf V}+{\bf V}\times(\nabla\times{\bf V})
-{\bf H}\times(\nabla\times{\bf H})+\beta\Theta{\bf e}_3
+\tau{\bf V}\times{\bf e}_3-\nabla P,\label{NS}\\
\partial{\bf H}/\partial t=\,&
\eta\nabla^2{\bf H}+\nabla\times({\bf V}\times{\bf H}),\label{MI}\\
\partial\Theta/\partial t=\,&
\kappa\nabla^2\Theta-({\bf V}\cdot\nabla)\Theta+\delta V_3,\label{HT}\\
&\nabla\cdot{\bf V}=\nabla\cdot{\bf H}=0.\label{so}\end{align}
Here $\bf V$ is the flow velocity, $\bf H$ magnetic field,
$\Theta=\vartheta-\vartheta_1+\delta x_3$ the difference between
the temperature, $\vartheta$, and the steady state linear profile,
$\vartheta_1$ and $\vartheta_2$ prescribed temperatures at the upper and lower
horizontal boundaries $x_3=0$ and 1, respectively,
$\delta=\vartheta_1-\vartheta_2>0$ and $\tau=\sqrt{\rm Ta}$.
The Navier-Stokes equation \rf{NS} involves buoyancy, Coriolis and
Lorentz forces. No external body forces, heat or electric sources are
present, i.e., there are no source terms in equations \rf{NS}--\rf{so};
in this case, we call convection free. Since the small-scale state
${\bf W}=({\bf V,H},\Theta)$ is supposed to be steady, the time derivatives
in the l.h.s.~of \rf{NS}--\rf{HT} vanish.

Equations \rf{NS}--\rf{so} are invariant with respect to the position
of the axis of rotation, i.e., the equations are the same for any vertical
axis of rotation. In an inertial (e.g., steady) coordinate system, a periodic
arrangement of rotating convective cells, with the pressure periodic
in horizontal directions, is impossible. Periodicity in $x_1$ and $x_2$
becomes possible after the pressure is corrected by $\tau^2\rho^2/8$,
where $\rho$ is the distance to the axis of rotation (see \cite{Cha});
this term represents the centripetal force preserving the arrangement of cells
(regarded as rigid bodies) rotating with constant angular velocity $\tau/2$.
Clearly, rotation about a vertical axis is compatible with the symmetry
about a vertical line; however, since the governing equations are invariant
with respect to the position of the axis of rotation, the axis of rotation
does not necessarily pass through the centre of a periodicity cell or coincides
with the axis about which the flow is symmetric.

We consider horizontal boundaries of the fluid layer that are stress-free,
ideally electrically conducting and kept at constant temperatures. This
translates into the following equations:
\begin{align}
\partial V_1/\partial x_3=\partial V_2/\partial x_3=V_3&=0,\label{bcv}\\
\partial H_1/\partial x_3=\partial H_2/\partial x_3=H_3&=0,\label{bch}\\
\Theta&=0.\label{bct}
\end{align}

We define the mean over the fast spatial variables, $\la\cdot\ra$,
and the fluctuating, $\lb\cdot\rb$, parts of a field~$f$:
$$\la f({\bf X},T,{\bf x},t)\ra\equiv{1\over{\mathscr P}^2}\int_{-{\mathscr P}/2}^{{\mathscr P}/2}
\int_{-{\mathscr P}/2}^{{\mathscr P}/2}\int_0^1f({\bf X},T,{\bf x},t)\D{\bf x},\qquad
\lb f\rb\equiv f-\la f\ra.$$
Same definitions hold for a vector field $\bf f$, and we also define
the mean of its horizontal part,
$\la{\bf f}\ra_h\equiv\la f_1\ra{\bf e}_1+\la f_2\ra{\bf e}_2$, and
the fluctuating part $\lb{\bf f}\rb_h\equiv{\bf f}-\la{\bf f}\ra_h$.
Later (see section~\ref{ker}) it will become clear why these averaging
procedures agree with the mathematical nature of the problem. We note that
the Coriolis force can be reexpressed as $\tau\lb{\bf V}\rb_h\times{\bf e}_3$,
since the constant horizontal force $\tau\la{\bf V}\ra_h\times{\bf e}_3$ is
a gradient of a linear function of horizontal coordinates that modifies
pressure.

Upon this modification, the operator of linearisation of equations
\rf{NS}--\rf{HT} around the steady state $\bf W$, which we denote by
$\l=(\l^v,\l^h,\l^\theta)$, takes the form
\begin{align*}
\l^v({\bf v},{\bf h},\theta)=&
-\partial{\bf v}/\partial t+\nu\nabla^2{\bf v}
+{\bf V}\times(\nabla\times{\bf v})+{\bf v}\times(\nabla\times{\bf V})\\
&-{\bf H}\times(\nabla\times{\bf h})-{\bf h}\times(\nabla\times{\bf H})
+\beta\theta{\bf e}_3+\tau\lb{\bf v}\rb_h\times{\bf e}_3-\nabla p,\\
\l^h({\bf v},{\bf h},\theta)=&-\partial{\bf h}/\partial t
+\eta\nabla^2{\bf h}+\nabla\times({\bf V}\times{\bf h}+{\bf v}\times{\bf H}),\\
\l^\theta({\bf v},{\bf h},\theta)=&-\partial\theta/\partial t
+\kappa\nabla^2\theta-({\bf V}\cdot\nabla)\theta-({\bf v}\cdot\nabla)\Theta
+\delta v_3.
\end{align*}

We consider a perturbation of the steady state $\bf W$, whose amplitude
is O($\varepsilon$). The perturbed state, $({\bf V}+\varepsilon{\bf v},\,
{\bf H}+\varepsilon{\bf h},\,\Theta+\varepsilon\theta)$, satisfies equations
\rf{NS}--\rf{so}, whereby
\begin{align}
\l^v({\bf v},{\bf h},\theta)=&
-\varepsilon({\bf v}\times(\nabla\times{\bf v})
-{\bf h}\times(\nabla\times{\bf h}))+\nabla p,\label{pNS}\\
\l^h({\bf v},{\bf h},\theta)=&
-\varepsilon\nabla\times({\bf v}\times{\bf h}),\label{pMI}\\
\l^\theta({\bf v},{\bf h},\theta)=&
\,\varepsilon({\bf v}\cdot\nabla)\theta,\label{pHT}\\
\nabla\cdot{\bf v}=&\,\nabla\cdot{\bf h}=0.\label{pso}\end{align}

Following the recipes of the homogenisation techniques, we now assume that
the perturbations depend on the slow horizontal spatial variables
${\bf X}=\varepsilon(x_1,x_2)$ and on the slow time, $T=\varepsilon^st$, where
$s\ge 1$, and expand the perturbation in a power series in the small
scale ratio $\varepsilon$:
\begin{align}
{\bf v}=&\sum_{n=0}^\infty{\bf v}_n({\bf X},T,{\bf x},t)\varepsilon^n,\label{vs}\\
{\bf h}=&\sum_{n=0}^\infty{\bf h}_n({\bf X},T,{\bf x},t)\varepsilon^n,\label{hs}\\
\theta=&\sum_{n=0}^\infty\theta_n({\bf X},T,{\bf x},t)\varepsilon^n.\label{ts}
\end{align}
Upon substituting these series into the governing equations \rf{pNS}--\rf{pso},
we obtain a hierarchy of systems of equations. In principle, it can be solved
at any order, and thus complete series \rf{vs}--\rf{ts} can be reconstructed.
In this paper we are only interested in the leading-order terms and therefore
will only consider the first 3 systems in the hierarchy.

The solenoidality conditions \rf{pso} at order $\varepsilon^n$ yield
equations, whose mean and fluctuating parts for any $n\ge0$ are
\begin{align}
\nabla_{\bf X}\cdot\la{\bf v}_n\ra_h=0,\label{dmv}\\
\nabla_{\bf X}\cdot\la{\bf h}_n\ra_h=0,\label{dmh}\\
\nabla_{\bf x}\cdot{\bf v}_n+\nabla_{\bf X}\cdot\lb{\bf v}_{n-1}\rb_h&=0,\label{dov}\\
\nabla_{\bf x}\cdot{\bf h}_n+\nabla_{\bf X}\cdot\lb{\bf h}_{n-1}\rb_h&=0\label{doh}
\end{align}
(any term of expansions with a negative index is zero by definition). We can
now comment on the modified form of the Coriolis force term in the operator
$\l^v$ in \rf{pNS}: By virtue of solenoidality \rf{dmv} of the mean field,
we can introduce a stream function, $\psi$, whereby
$\la{\bf v}\ra_h=(-\partial\psi/\partial X_2,\,\partial\psi/\partial X_1,\,0)$.
Thus, the difference between the original, $\tau{\bf v}\times{\bf e}_3$,
and modified, $\tau\lb{\bf v}\rb_h\times{\bf e}_3$, forms is equal to
$\tau\la{\bf v}\ra_h\times{\bf e}_3=\tau\nabla_{\bf X}\psi$, which
is compensated by modifying the pressure perturbation, $p$. So, although now
the flow perturbation, $\bf v$, depends on the slow spatial variables,
the operator of linearisation, $\l^v$, of the Navier--Stokes equation
\rf{NS} can still be used in the form \rf{pNS}.

We assume henceforth that the pressure perturbation $p$ is $\mathscr P$-periodic
in the fast horizontal variables.

\subsection{Order $\varepsilon^0$ equations}\label{ord0}

This section is devoted to the study of the generalised kernel of the operator
of linearisation $\l$. We find that the operator acting on the generalised
kernel has two Jordan normal form blocks of size two, and, consequently,
the means of the flow and magnetic components of neutral modes of $\l$,
${\bf S}_k$ for $k=1,2$, are related by a matrix $\U$ of size two introduced
below. The perturbed states are missing the symmetries which make this matrix
vanish. As a result, it will turn out that because of the solenoidality
in the slow variable of the spatially averaged perturbations of the form
\rf{vs}--\rf{ts}, the latter can be described by a self-consistent system
of amplitude equations only if they depend on a single spatial variable
$Y={\bf q}\cdot\bf X$, where a unit vector $\bf q$ is normal to an eigenvector
of $\U$.

\subsubsection{The Jordan normal form of the operator of linearisation\\
in the invariant subspace associated with the zero eigenvalue}\label{ker}

The first system of equations in the hierarchy is obtained at order
$\varepsilon^0$:
\BE\l({\bf v}_0,{\bf b}_0,\theta_0)=0,\qquad
\nabla_{\bf x}\cdot{\bf v}_0=\nabla_{\bf x}\cdot{\bf b}_0=0,\EE{fap}
i.e., the field $({\bf v}_0,{\bf b}_0,\theta_0)$ belongs to the kernel
of the operator of linearisation, $\l$; it must satisfy the requirements
for perturbation, i.e., the solenoidality conditions for $\bf v$ and $\bf h$,
$\mathscr P$-periodicity in the fast horizontal variables
and boundary conditions \rf{bcv}--\rf{bct}.

As usual, we derive the adjoint operator $\l^*=(\l^{*v},\l^{*h},\l^{*\theta})$
in the Lebesgue space L$_2$ of solenoidal vector fields by using vector analysis
identities and, essentially, integrating by parts the l.h.s.~of the defining
equation
\BE\l{\bf w}\cdot{\bf w}^*={\bf w}\cdot\l^*{\bf w}^*:\EE{defa}
\begin{align*}
\l^{*v}({\bf v},{\bf h},\theta)=&\,\partial{\bf v}/\partial t
+\nu\nabla^2{\bf v}-\nabla\times\left({\bf V}\times{\bf v}\right)\\
&+\p({\bf H}\times(\nabla\times{\bf h})
-{\bf v}\times(\nabla\times{\bf V})+\delta\theta{\bf e}_3
-\tau\lb{\bf v}\rb\times{\bf e}_3+\Theta\nabla\theta),\\
\l^{*h}({\bf v},{\bf h},\theta)=&\,\partial{\bf h}/\partial t
+\eta\nabla^2{\bf h}+\nabla\times({\bf H}\times{\bf v})
+\p({\bf v}\times(\nabla\times{\bf H})
-{\bf V}\times(\nabla\times{\bf h})),\\
\l^{*\theta}({\bf v},{\bf h},\theta)=&\,\partial\theta/\partial t
+\kappa\nabla^2\theta+({\bf V}\cdot\nabla)\theta+\beta v_3.
\end{align*}
Here $\p$ is the projection onto the subspace of solenoidal vector fields
satisfying the requirements for perturbation.

We assume henceforth that the small-scale perturbed steady state
${\bf W}=({\bf V,H},\Theta)$ is stable to small-scale perturbations, i.e.,
real parts of all eigenvalues of $\l$ are non-positive. It is then natural
to assume that the evolution of large-scale perturbations of $\bf W$ is examined
after all the fast-time transients have decayed. Since the perturbed state is
steady, we will assume that the perturbations do not depend on the fast time,
and hence the derivatives in $t$ in the definitions of the operators $\l$ and
$\l^*$ will be omitted. (It would be necessary to preserve them
when investigating large-scale stability of time-periodic states.)

Since $\l^*({\bf e}_k,0,0)=\l^*(0,{\bf e}_k,0)=0$ for $k=1,2$,
the kernel of the operator $\l$ is at least four-dimensional.
Differentiating \rf{NS}--\rf{so} in $x_k$ for $k=1,2$, we find that
\BE{\bf S}_{k+2}=\partial{\bf W}/\partial x_k\EE{s34}
belong to the kernel of $\l$. Furthermore,
\BE\l({\bf e}_k,0,0)=-{\bf S}_{k+2}\in\ker\l;\EE{Lek}
therefore, the Jordan normal form of $\l$ involves two blocks of size 2
associated with the eigenvalue zero, and $\l$ has at least 6 eigenfields
associated with this eigenvalue, at least two of which are generalised.
Henceforth, we assume the generic case where zero is a 6-fold (generalised)
eigenvalue of $\l$. Clearly, $\la\l^v({\bf w})\ra_h=\la\l^h({\bf w})\ra_h=0$
for any ${\bf w}=({\bf v},{\bf h},\theta)$ satisfying the requirements
for perturbation. Consequently, it is simple to show that any eigenfield,
a generalised one or not, which has a non-zero mean horizontal part
of the flow and/or magnetic component, is associated with the zero
eigenvalue. Such eigenfields exist, because eigenfields of the elliptic
operator $\l$ constitute a complete basis in the Lebesgue space L$_2$;
hence, the kernel of $\l$ involves the fields
${\bf S}_k=({\bf S}_k^v,{\bf S}_k^h,\,S_k^\theta)$
for $k=1,2$, satisfying the requirements for perturbation and such that
$$\l{\bf S}_k=0,\qquad
{\bf S}_k^v=\lb{\bf S}_k^v\rb_h+\sum_{m=1}^2u_{mk}{\bf e}_m,\qquad
{\bf S}_k^h=\lb{\bf S}_k^h\rb_h+{\bf e}_k.$$
The {\bf first auxiliary problem} is to find the fields ${\bf S}_k$
and the matrix $\displaystyle\U=\left[
\begin{matrix}u_{11}&\!\!u_{12}\\u_{21}&\!\!u_{22}\end{matrix}\right]$.
Since normal Jordan forms of an operator and its adjoint coincide,
there exist two generalised eigenfields
${\bf S}_n^*=({\bf S}_n^{*v},{\bf S}_n^{*h},S_n^{*\theta})$, $n=1,2$,
associated with the eigenvalue zero of $\l^*$, i.e., such that
\BE\l^*{\bf S}_n^*\!=\!\left(\sum_{m=1}^2u'_{mn}{\bf e}_m,\,{\bf e}_n,\,0\!\right),
\quad\la{\bf S}^{*v}_n\ra_h\!=\!\la{\bf S}^{*h}_n\ra_h\!=0,\quad
\nabla_{\bf x}\!\cdot\!{\bf S}^{*v}_n\!=\!\nabla_{\bf x}\!\cdot\!{\bf S}^{*h}_n\!=0
\EE{Sas}
and also satisfying the requirements for perturbation. It is straightforward
to demonstrate that
\BE\left[\begin{matrix}u'_{11}&u'_{12}\\u'_{21}&u'_{22}\end{matrix}\right]
=-(\U^t)^{-1},\EE{Up}
where the superscript $t$ denotes the transpose of a matrix.

By linearity of the system of order $\varepsilon^0$ equations \rf{fap}, its
solution is
\BE{\bf w}_0=({\bf v}_0,{\bf h}_0,\theta_0)
=\sum_{k=1}^4c_{0k}({\bf X},T)\,{\bf S}_k({\bf x}),\qquad
p_0=\sum_{k=1}^4c_{0k}({\bf X},T)\,S^p_k({\bf x}),\EE{w0}
where $p_0$ and $S^p_k$ denote the gradient potentials arising
in the equations $\l^v{\bf w}_0=0$ and $\l^v{\bf S}_k=0$,
respectively. The horizontal mean of the magnetic component of the first
relation in \rf{w0} reduces to $c_{0k}=\la h_{0,k}\ra$ for $k=1,2$,
and hence $\la{\bf v}_0\ra_h=\U\la{\bf h}_0\ra_h$.

The solvability condition for a problem
$\l{\bf w}=({\bf f}^v,{\bf f}^h,f^\theta)$ consists of orthogonality
of the r.h.s.~to the kernel of the adjoint $\l^*$. Since the kernel of $\l^*$
is comprised of constant vectors in the horizontal flow and magnetic
components, the solvability condition reduces to
$\la{\bf f}^v\ra_h=\la{\bf f}^h\ra_h=0$. It is straightforward to establish
that this condition is necessary; that it is sufficient follows from
the Fredholm alternative theorem applied to the problem
$(\nabla^2)^{-1}\l{\bf w}=(\nabla^2)^{-1}({\bf f}^v,{\bf f}^h,f^\theta)$
(the proof is essentially the same as in section~3.2 in \cite{vz}). It is due
to this form of the solvability condition that spatial averaging
over the periodicity domain is the only appropriate kind of averaging
in the problem under investigation.

\subsubsection{Solenoidality of the mean fields}\label{slosol}

Here we discuss how to satisfy the solenoidality conditions \rf{dmv}--\rf{dmh}.
We assume that amplitudes $c_{0k}$ are space-periodic in the slow variables
${\bf X}=(X_1,\,X_2)$ and expand the amplitudes in Fourier series, which
it is convenient to express in the form
\BE\la{\bf v}_0\ra_h=\sum_{k,\,|\bf q|=1}(v_{1,k\bf q}{\bf e}_1
+v_{2,k\bf q}{\bf e}_2)\,\e^{\i k\bf q\cdot X},\quad
\la{\bf h}_0\ra_h=\sum_{k,\,|\bf q|=1}(h_{1,k\bf q}{\bf e}_1
+h_{2,k\bf q}{\bf e}_2)\,\e^{\i k\bf q\cdot X},\EE{Fou}
and consider the conditions independently for each unit wave vector
${\bf q}=(q_1,q_2)$.

Solenoidality of $\la{\bf h}_0\ra_h$ then requires
$(h_{1,k\bf q},h_{2,k\bf q})=C_{k\bf q}{\bf q}^\perp$, where
${\bf q}^\perp=(q_2,-q_1)$, and hence solenoidality of
$\displaystyle\left[\!\!\begin{array}{c}v_{1,k\bf q}\\v_{2,k\bf q}\end{array}\!\!\right]\!\!
=\U\left[\!\!\begin{array}{c}h_{1,k\bf q}\\h_{2,k\bf q}\end{array}\!\!\right]$
is equivalent to the orthogonality condition $\U{\bf q}^\perp\cdot{\bf q}=0$.
It implies ${\bf q}^\perp\parallel\U{\bf q}^\perp$; thus,
${\bf q}^\perp$ is an eigenvector of $\U$. We denote the associated
eigenvalue by $\lambda={\bf q}^\perp\cdot\U{\bf q}^\perp$.
In the basis $\{{\bf q}^\perp,\bf q\,\}$, the matrix $\U$ is upper triangular;
consequently, the wave vector $\bf q$ is an eigenvector of $\U^t$
associated with the second eigenvalue $\lambda'={\bf q}\cdot\U\bf q$.

Thus, by virtue of the solenoidality conditions
the series \rf{Fou} may involve only two wave vectors, ${\bf q}^1$ and
${\bf q}^2$, that are normal to eigenvectors of $\U$; hence $\la{\bf v}_0\ra_h$
and $\la{\bf h}_0\ra_h$ split into sums of vector fields, each depending on just
one spatial variable, $Y_1={\bf q}^1\cdot\bf X$ or $Y_2={\bf q}^2\cdot\bf X$.
We will see in section~\ref{ord2} that the system of amplitude equations for
$c_{0k}$ is nonlinear (cubic). Therefore, a weakly nonlinear perturbation
depends on one spatial variable $Y={\bf q}\cdot\bf X$ such that ${\bf q}^\perp$
is an eigenvector of $\U$ (otherwise, products of functions of $Y_1$ and $Y_2$
are generated by the nonlinearity of the governing equations for $c_{0k}$).
Perturbations with initial amplitudes that violate this one-dimensionality
restriction cannot thus be expanded in the series \rf{vs}--\rf{ts} and are not
described by homogenised equations that we will derive. Arising for $\U\ne0$,
this is the main difference with the cases considered in \cite{vz}. Summarising,
\BE\left[\!\!\!\begin{array}{c}\la v_{0,1}\ra\\\la v_{0,2}\ra\end{array}\!\!\!\right]
=\lambda C(Y,T){\bf q}^\perp,~~~~
\left[\!\!\!\begin{array}{c}\la h_{0,1}\ra\\\la h_{0,2}\ra\end{array}\!\!\!\right]
=C(Y,T){\bf q}^\perp,~~~~
\U{\bf q}^\perp=\lambda{\bf q}^\perp,~~~~Y={\bf q}\cdot{\bf X}.~~\EE{v0h0}

We will henceforth use the same notation for the two-dimensional vector
$\bf q$ and the three-dimensional horizontal vector, whose horizontal component
coincides with $\bf q$; similarly for ${\bf q}^\perp$.

\subsection{Order $\varepsilon^1$ equations}\label{ord1}

At order $\varepsilon^1$ we obtain the system of equations
\begin{align}
\l^v({\bf v}_1,{\bf h}_1,\theta_1)
&+2\nu(\nabla_{\bf x}\cdot\nabla_{\bf X}){\bf v}_0
+{\bf V}\times(\nabla_{\bf X}\times{\bf v}_0)
-{\bf H}\times(\nabla_{\bf X}\times{\bf h}_0)\nonumber\\
&+{\bf v}_0\times(\nabla_{\bf x}\times{\bf v}_0)
-{\bf h}_0\times(\nabla_{\bf x}\times{\bf h}_0)-\nabla_{\bf X}\,p_0=0,\label{vo1}\\
\l^h({\bf v}_1,{\bf h}_1,\theta_1)
&+2\eta(\nabla_{\bf x}\cdot\nabla_{\bf X}){\bf h}_0
+\nabla_{\bf X}\times({\bf v}_0\times{\bf H}
+{\bf V}\times{\bf h}_0)+\nabla_{\bf x}\times({\bf v}_0\times{\bf h}_0)=0,\label{ho1}\\
\l^\theta({\bf v}_1,{\bf h}_1,\theta_1)
&+2\kappa(\nabla_{\bf x}\cdot\nabla_{\bf X})\theta_0
-({\bf V}\cdot\nabla_{\bf X})\theta_0-({\bf v}_0\cdot\nabla_{\bf x})\theta_0=0\label{to1}.
\end{align}
For the slow time $T=\varepsilon t$, these equations would also involve
the derivatives of the respective fields in the slow time. The solvability
condition then defines the operator of the significant combined $\alpha$-effect:
it consists of spatial means of linear (in ${\bf w}_0$) terms
in the horizontal components of the flow and magnetic equations. However,
the perturbed state ${\bf W}=({\bf V},{\bf H},\Theta)$ is supposed to be
symmetric about the vertical axis, $x_3$. Consequently, solutions
to the first auxiliary problems are antisymmetric, the r.h.s.~of the equations
are symmetric, and the $\alpha$-effect operator is zero: convective
hydromagnetic states with such a symmetry lack any significant (i.e.,
responsible for the evolution of large-scale amplitudes) $\alpha$-effect.
In the absence of significant $\alpha$-effect, we have to switch to the slow
time $T=\varepsilon^2t$. Since the solvability condition is verified,
a solution to this system exists and, by linearity, takes the form
\BE\left(\begin{array}{c}{\bf v}_1\\{\bf h}_1\\\theta_1\end{array}\right)
=\sum_{k=1}^4\left(c_{1k}{\bf S}_k+\sum_{m=1}^2{\bf G}_{mk}
{\partial c_{0k}\over\partial X_m}+\sum_{m=1}^4{\bf Q}_{mk}c_{0k}c_{0m}\right),\EE{pe1}
where ${\bf G}_{mk}({\bf x},t)$ for $k=1,...,4$ and $m=1,2$ are solutions
to the {\bf second auxiliary problem}:
\begin{align}
\l^v{\bf G}_{mk}=&-2\nu\partial{\bf S}_k^v/\partial x_m
-{\bf V}\times({\bf e}_m\times{\bf S}_k^v)
+{\bf H}\times({\bf e}_m\times{\bf S}_k^h)+S_k^p{\bf e}_m,\label{gv}\\
\l^h{\bf G}_{mk}=&-2\eta\partial{\bf S}_k^h/\partial x_m
-{\bf e}_m\times({\bf V}\times{\bf S}_k^h+{\bf S}_k^v\times{\bf H}),\label{gh}\\
\l^\theta{\bf G}_{mk}=&-2\kappa\partial S_k^\theta/\partial x_m
+V_m S_k^\theta,\label{gt}\\
&\nabla\cdot{\bf G}_{mk}^v=-\lb S_{k,m}^v\rb,\qquad
\nabla\cdot{\bf G}_{mk}^h=-\lb S_{k,m}^h\rb\label{gs}
\end{align}
and ${\bf Q}_{mk}({\bf x},t)$ for $m,k=1,...,4$ are solutions
to the {\bf third auxiliary problem}:
\begin{align}
2\l^v{\bf Q}_{mk}=&-{\bf S}_k^v\!\times\!(\nabla\!\times\!{\bf S}_m^v)
+{\bf S}_k^h\!\times\!(\nabla\!\times\!{\bf S}_m^h)
-{\bf S}_m^v\!\times\!(\nabla\!\times\!{\bf S}_k^v)
+{\bf S}_m^h\!\times\!(\nabla\!\times\!{\bf S}_k^h),\label{qv}\\
2\l^h{\bf Q}_{mk}=&-\nabla\!\times\!({\bf S}_k^v\!\times\!{\bf S}_m^h
+{\bf S}_m^v\!\times\!{\bf S}_k^h),\label{qh}\\
2\l^\theta{\bf Q}_{mk}=&\,({\bf S}_k^v\cdot\nabla)S_m^\theta
+({\bf S}_m^v\cdot\nabla)S_k^\theta\label{qt},\\
\nabla\cdot{\bf Q}_{mk}^v=&\,\nabla\cdot{\bf Q}_{mk}^h=0\label{qs}
\end{align}
(for the symmetrised last sum in \rf{pe1}).
The potential $p_1$ in \rf{vo1} can be expressed in terms of the gradient
potentials $G^p_k$ and $Q^p_k$ arising in \rf{gv} and \rf{qv},
where the flow components ${\bf G}_{mk}^v$ and ${\bf Q}_{mk}^v$ satisfy
relations \rf{gs} and \rf{qs}, respectively:
$$p_1=\sum_{k=1}^4\left(c_{1k}S^p_k+\sum_{m=1}^2G^p_{mk}{\partial c_{0k}\over\partial X_m}
+\sum_{m=1}^4Q^p_{mk}c_{0k}c_{0m}\right),\qquad\la p_1\ra=0.$$

For the same symmetry reasons as for the full equations \rf{vo1}--\rf{to1},
the r.h.s.~of the equations in formulations of the second and third auxiliary
problems are symmetric, and hence the solvability conditions for them
are satisfied. Symmetric and antisymmetric fields constitute invariant
subspaces of the operator of linearisation $\l$, and thus ${\bf G}_{mk}$ and
${\bf Q}_{mk}$ are symmetric about the vertical axis.

Examining the system of equations for the difference
${\bf G}_{14}-{\bf G}_{23}$, it is straightforward (albeit tedious) to show that
\BE{\bf G}_{14}-{\bf G}_{23}=({\bf V}\times{\bf e}_3,{\bf H}\times{\bf e}_3,0).
\EE{gmk}

Differentiating the equation $\l({\bf S}_k)=0$ in $x_1$ and $x_2$ and
comparing the results with \rf{qv}--\rf{qs}, we find
\BE\partial{\bf S}_k/\partial x_{m-2}=2{\bf Q}_{mk}=2{\bf Q}_{km},\EE{qmk}
for any $k=1,...,4$ and $m=3,4$.

\subsection{Order $\varepsilon^2$ equations}\label{ord2}

At order $\varepsilon^2$ we obtain the system of equations
\begin{flalign}
\partial{\bf v}_0/\partial T=&\,\l^v({\bf v}_2,{\bf h}_2,\theta_2)
+2\nu(\nabla_{\bf x}\cdot\nabla_{\bf X}){\bf v}_1+\nu\nabla_{\bf X}^2{\bf v}_0\nonumber\\
&\!+{\bf V}\times(\nabla_{\bf X}\times{\bf v}_1)
-{\bf H}\times(\nabla_{\bf X}\times{\bf h}_1)
+{\bf v}_0\times(\nabla_{\bf X}\times{\bf v}_0
+\nabla_{\bf x}\times{\bf v}_1)\label{vo2}\\
&\!-{\bf h}_0\times(\nabla_{\bf X}\times{\bf h}_0+\nabla_{\bf x}\times{\bf h}_1)
+{\bf v}_1\times(\nabla_{\bf x}\times{\bf v}_0)
-{\bf h}_1\times(\nabla_{\bf x}\times{\bf h}_0)-\nabla_{\bf X}p_1,\nonumber\\
\partial{\bf h}_0/\partial T=&\,\l^h({\bf v}_2,{\bf h}_2,\theta_2)
+2\eta(\nabla_{\bf x}\cdot\nabla_{\bf X}){\bf h}_1
+\eta\nabla_{\bf X}^2{\bf h}_0\label{ho2}\\
&\!+\nabla_{\bf X}\times({\bf v}_1\times{\bf H}+{\bf V}\times{\bf h}_1
+{\bf v}_0\times{\bf h}_0)+\nabla_{\bf x}\times({\bf v}_0\times{\bf h}_1
+{\bf v}_1\times{\bf h}_0),\nonumber\\
\partial\theta_0/\partial T=&\,\l^\theta({\bf v}_2,{\bf h}_2,\theta_2)
+2\kappa(\nabla_{\bf x}\cdot\nabla_{\bf X})\theta_1
+\kappa\nabla_{\bf X}^2\theta_0\label{to2}\\
&\!-({\bf v}_1\cdot\nabla_{\bf x})\theta_0
-({\bf v}_0\cdot\nabla_{\bf x})\theta_1-({\bf v}_0\cdot\nabla_{\bf X})\theta_0
-({\bf V}\cdot\nabla_{\bf X})\theta_1.\nonumber
\end{flalign}

Two equations for the evolution of amplitudes of large-scale perturbations are
the solvability conditions for the system \rf{vo2}--\rf{to2}: orthogonality to
the kernel of the adjoint operator, $\l^*$, i.e., the requirement of vanishing
of the horizontal means of the flow and magnetic equations. By~\rf{vo2},
\BE{\partial\la{\bf v}_0\ra_h\over\partial T}=\nu\nabla_{\bf X}^2\la{\bf v}_0\ra_h
+\sum_{j=1}^2\sum_{k=1}^4\!\left(\sum_{m=1}^2{\bf D}_{jmk}^v
{\partial^2c_{0k}\over\partial X_j\partial X_m}
+\!\sum_{m=1}^4{\bf A}_{jmk}^v{\partial(c_{0m}c_{0k})\over\partial X_j}\!\right)
-\nabla_{\bf X}\la\widetilde{p}_1\ra_h,\EE{veq}
where we have denoted
$${\bf D}_{jmk}^v=\la{\bf V}\times({\bf e}_j\times{\bf G}_{mk}^v)
-{\bf H}\times({\bf e}_j\times{\bf G}_{mk}^h)\ra_h,$$
$${\bf A}_{jmk}^v=\la{\bf V}\times({\bf e}_j\times{\bf Q}_{mk}^v)
-{\bf H}\times({\bf e}_j\times{\bf Q}_{mk}^h)-{\bf S}_k^vS_{m,j}^v
+{\bf S}_k^hS_{m,j}^h\ra_h,$$
and from \rf{ho2},
\BE{\partial\la{\bf h}_0\ra_h\over\partial T}
=\eta\nabla_{\bf X}^2\la{\bf h}_0\ra_h-{\bf e}_3\times\nabla_{\bf X}
\sum_{k=1}^4\left(\sum_{m=1}^2D_{mk}^h{\partial c_{0k}\over\partial X_m}
+\sum_{m=1}^4A_{mk}^hc_{0m}c_{0k}\right),\EE{heq}
where
$$D_{mk}^h=\la({\bf V}\times{\bf G}_{mk}^h-{\bf H}\times{\bf G}_{mk}^v)
\cdot{\bf e}_3\ra,$$
$$A_{mk}^h=\la({\bf V}\times{\bf Q}_{mk}^h-{\bf H}\times{\bf Q}_{mk}^v
+{\bf S}_m^v\times{\bf S}_k^h)\cdot{\bf e}_3\ra.$$
The quantities ${\bf D}_{jmk}^v$ and $D_{mk}^h$ describe eddy diffusivity,
and ${\bf A}_{jmk}^v$ and $A_{mk}^h$ eddy advection.

Significant simplifications stem from relations \rf{v0h0}. The last term
in \rf{veq} serves to eliminate from this equation the component, parallel
to $\bf q$; consequently, \rf{veq} is equivalent to the scalar equation
\BE\lambda{\partial C\over\partial T}=\nu\lambda{\partial^2C\over\partial Y^2}
+\sum_{j=1}^2\sum_{k=1}^4\left(\sum_{m=1}^2
{\bf q}^\perp\cdot{\bf D}_{jmk}^vq_jq_m{\partial^2c_{0k}\over\partial Y^2}
+\sum_{m=1}^4{\bf q}^\perp\cdot{\bf A}_{jmk}^vq_j
{\partial(c_{0m}c_{0k})\over\partial Y}\right).\EE{vsc}
All terms in \rf{heq} are parallel to ${\bf q}^\perp$,
and, in view of \rf{qmk}, it reduces to the scalar equation
\BE{\partial C\over\partial T}={\cal D}^h{\partial^2C\over\partial Y^2}
+\sum_{k=3}^4{\cal D}^h_k{\partial^2c_{0k}\over\partial Y^2}
+{\cal A}^h{\partial(C^2)\over\partial Y},\EE{hsc}
where
$${\cal A}^h=\sum_{k=1}^2\sum_{m=1}^2(-1)^{m+k}A_{mk}^hq_{3-m}q_{3-k},$$
$${\cal D}^h=\eta-\sum_{k=1}^2\sum_{m=1}^2(-1)^kD_{mk}^hq_mq_{3-k},\qquad
{\cal D}^h_k=\sum_{m=1}^2D_{mk}^hq_m.$$

Relations \rf{gmk} imply the following identities for eddy coefficients
in \rf{veq}--\rf{heq}:
$$D_{14}^h=D_{23}^h,\qquad{\bf D}_{j14}^v-{\bf D}_{j23}^v=
\la-({\bf V}\times{\bf e}_3)V_j+({\bf H}\times{\bf e}_3)H_j\ra_h$$
for $j=1,2$. Expressions for coefficients in the last sum in \rf{vsc} can be
simplified by using an identity that follows from \rf{qmk}:
$$\sum_j{\bf q}^\perp\cdot({\bf A}_{jmk}^v+{\bf A}_{jkm}^v)q_j=\sum_j
{\bf q}^\perp\cdot\la{\bf S}_m^hS_{k,j}^h-{\bf S}_m^vS_{k,j}^v\ra_hq_j,$$
where $m=3,4$ or $k=3,4$.

\pagebreak
Since the two equations \rf{vsc} and \rf{hsc} involve the slow time derivative
of the same quantity, $C$, a simpler non-evolutionary equation can be derived,
equivalent to any of the two. Multiplying \rf{hsc} by $\lambda$, subtracting
it from \rf{vsc} and integrating the result in $Y$ we find
\begin{gather}
{\partial\over\partial Y}\Bigg(\Big(\lambda(\nu-{\cal D}^h)
+\sum_{j=1}^2\sum_{m=1}^2\sum_{k=1}^2
{\bf q}^\perp\cdot{\bf D}_{jmk}^vq_jq_m(-1)^{k+1}q_{3-k}\Big)C\label{nonev}\\
+\sum_{k=3}^4\Big(-\lambda{\cal D}^h_k+\sum_{j=1}^2\sum_{m=1}^2
{\bf q}^\perp\cdot{\bf D}_{jmk}^vq_jq_m\Big)c_{0k}\Bigg)
=\lambda{\cal A}^hC^2-\sum_{j=1}^2\sum_{k=1}^4\sum_{m=1}^4
{\bf q}^\perp\cdot{\bf A}_{jmk}^vq_jc_{0m}c_{0k}+K(T),\nonumber
\end{gather}
where $K(T)$ is a function of slow time alone. We will consider perturbations
(in particular, $c_{0k}$) that are periodic in $Y$; by
appropriately rescaling $\varepsilon$, the period then can be made equal
to $2\pi$ (in agreement with \rf{Fou}). Averaging \rf{nonev} in $Y$ over
a period uniquely determines $K(T)$.

\subsubsection{The closing evolutionary equation}\label{add}

Another evolutionary equation will be derived from \rf{vo2}--\rf{to2} owing
to the fact that the generalised kernels of the operators $\l$ and $\l^*$
are six-dimensional. We begin by making a general remark on the means
of the perturbation fields, which will be used in the derivations.
Let $\LA\cdot\RA$ denote the averaging in $Y$ over a period:
$$\LA f(Y,T)\RA={1\over2\pi}\int_0^{2\pi}f(Y,T)\D Y.$$
Clearly, application of the two averagings, $\la\cdot\ra$ and $\LA\cdot\RA$,
to the flow and magnetic equations obtained at order $\varepsilon^{n+2}$ yields
$\partial\LA\la{\bf v}_n\ra\RA/\partial T=0$ (there is no contribution
from nonlinear terms by virtue of \rf{dmv}--\rf{doh}) and
$\partial\LA\la{\bf h}_n\ra\RA/\partial T=0$. Thus, the means over both
slow and fast variables of each term of the expansions of the flow and magnetic
perturbation are independent of time. We investigate instabilities (if any)
of the steady state $\bf W$ developing in their own right and not due to a mean
flow through the layer, or an imposed mean magnetic field, and therefore demand
\BE\LA\la{\bf v}_n\ra\RA=\LA\la{\bf h}_n\ra\RA=0.\EE{means}

We split ${\bf v}_2$ and ${\bf h}_2$, into solenoidal and gradient parts:
\begin{align}
{\bf v}_2={\bf v}_2^{\rm sol}+\nabla_{\bf x}v_2^{\rm gr},\qquad
&\nabla_{\bf x}\cdot{\bf v}_2^{\rm sol}=0,\qquad
\nabla^2_{\bf x}v_2^{\rm gr}=-\nabla_{\bf X}\cdot{\bf v}_1,\label{vspl} \\
{\bf h}_2={\bf h}_2^{\rm sol}+\nabla_{\bf x}h_2^{\rm gr},\qquad
&\nabla_{\bf x}\cdot{\bf h}_2^{\rm sol}=0,\qquad
\nabla^2_{\bf x}h_2^{\rm gr}=-\nabla_{\bf X}\cdot{\bf h}_1,\label{hspl}
\end{align}
where we have used relations \rf{dov} and \rf{doh} for $n=2$.
Upon substituting these sums into \rf{vo2}--\rf{to2}, we recast this system as
\BE\l{\bf w}^{\rm sol}_2=\partial{\bf w}_0/\partial T-{\bf F},\EE{ws2}
where ${\bf w}^{\rm sol}_2=({\bf v}^{\rm sol}_2,{\bf h}^{\rm sol}_2,\theta_2)$
and $\bf F$ denotes the sum of all the remaining terms in these equations.
Scalar multiplying \rf{ws2} by ${\bf S}_j^*$, $j=1,2$, we find using \rf{Sas}:
$${\partial\over\partial T}\la{\bf w}_0\cdot{\bf S}_j^*\ra
-\la{\bf F}\cdot{\bf S}_j^*\ra=\la\l{\bf w}^{\rm sol}_2\cdot{\bf S}_j^*\ra
=\la{\bf w}^{\rm sol}_2\cdot\l^*{\bf S}_j^*\ra
=\sum_{m=1}^2u'_{mj}\la v_{2,m}\ra+\la h_{2,j}\ra$$
(note that, by our constructions in section~\ref{ker}, relation \rf{defa}
defining the adjoint operator $\l^*$ is only valid for fields $\bf w$ and
$\bf w^*$, whose flow and magnetic field components are solenoidal in the
fast spatial variables --- this has forced us to perform the Helmholtz
decompositions \rf{vspl}--\rf{hspl}). Now, by virtue of \rf{Up},
\BE\sum_{j=1}^2\left({\partial\over\partial T}\la{\bf w}_0\cdot{\bf S}_j^*\ra
-\la{\bf F}\cdot{\bf S}_j^*\ra\right){\bf e}_j=-\U^{-1}\la{\bf v}_2\ra_h
+\la{\bf h}_2\ra_h.\EE{Oh}

\pagebreak\noindent
However, by \rf{dmv} and \rf{dmh} for $j=2$, $\la{\bf v}_2\ra_h$ and
$\la{\bf h}_2\ra_h$ are solenoidal in the slow variable $\bf X$, and, since
their spatial dependence is exclusively on $Y=\bf X\cdot q$, they are parallel
to ${\bf q}^\perp$ (Fourier expansion in $Y$ can be used to show this
for zero-mean fields; see \rf{means} for $n=2$). Furthermore,
since ${\bf q}^\perp$ is an eigenvector of $\U$, the r.h.s.~of \rf{Oh}
is parallel to ${\bf q}^\perp$. Thus, scalar multiplying \rf{Oh} by $\bf q$,
we obtain a closed equation in $c_{0k}$:
\BE\sum_{j=1}^2\left(\sum_{k=1}^4{\partial c_{0k}\over\partial T}
\la{\bf S}_k\cdot{\bf S}_j^*\ra-\la{\bf F}\cdot{\bf S}_j^*\ra\right)q_j=0.
\EE{miss}
It can be simplified using relations \rf{Lek} and \rf{Sas}:
\BE\la{\bf S}_{k+2}\cdot{\bf S}_j^*\ra
=-\la\l({\bf e}_k,0,0)\cdot{\bf S}_j^*\ra=-u'_{kj}.\EE{SkSj}
Since $\bf q$ is an eigenvector of the matrix \rf{Up},
$$\sum_{j=1}^2\la{\bf S}_{k+2}\cdot{\bf S}_j^*\ra q_j=-\sum_{j=1}^2 u'_{kj}q_j
={\bf e}_k\cdot(\U^t)^{-1}{\bf q}={q_k\over\lambda'}.$$
By virtue of \rf{v0h0}, \rf{miss} takes the form
\BE{\partial\over\partial T}\Big(\sum_{j=1}^2q_j\la(q_2{\bf S}_1
-q_1{\bf S}_2)\cdot{\bf S}_j^*\ra C+{1\over\lambda'}\sum_{k=3}^4q_{k-2}c_{0k}
\Big)=\sum_{j=1}^2q_j\la{\bf F}\cdot{\bf S}_j^*\ra.\EE{fslx}

It is straightforward to calculate the scalar products in the r.h.s.~of this
equation, using the expressions for the two leading terms in the expansion
of perturbation, that were found in sections~\ref{ker} and \ref{ord1}.
The result has a simple structure, but involves bulky coefficients:
\begin{gather}
\la{\bf F}\cdot{\bf S}_j^*\ra=\sum_{n=1}^2\sum_{m=1}^2\sum_{k=1}^4
{\partial^2c_{0k}\over\partial X_n\partial X_m}f_{nmkj}^1
+\sum_{n=1}^2\sum_{m=1}^4\sum_{k=1}^4{\partial c_{0k}\over\partial X_n}
c_{0m}f_{nmkj}^2\label{rhs}\\
+\sum_{n=1}^4\sum_{m=1}^4\sum_{k=1}^4c_{0n}c_{0m}c_{0k}f_{nmkj}^3,\nonumber
\end{gather}
where
\begin{align*}
f_{nmkj}^1=&\Big\la\Big(2\nu\,\partial{\bf G}_{mk}^v/\partial x_n
+\delta_m^n\nu{\bf S}_k^v+{\bf V}\times({\bf e}_n\times{\bf G}_{mk}^v)
-{\bf H}\times({\bf e}_n\times{\bf G}_{mk}^h)\Big)\cdot{\bf S}_j^{*v}\\
&-\,G_{mk}^pS_{j,n}^{*v}+\Big(2\eta\,\partial{\bf G}_{mk}^h/\partial x_n
+\delta_m^n\eta{\bf S}_k^h+{\bf e}_n\times({\bf V}\times{\bf G}_{mk}^h
-{\bf H}\times{\bf G}_{mk}^v)\Big)\cdot{\bf S}_j^{*h}\\
&+\Big(2\kappa\,\partial G_{mk}^\theta/\partial x_n
+\delta_m^n\kappa S_k^\theta-V_nG_{mk}^\theta\Big)S_j^{*\theta}
-\l(\nabla\widehat G_{mkn}^v,\nabla\widehat G_{mkn}^h,0)\cdot{\bf S}_j^*\Big\ra,\\
f_{nmkj}^2=&\Big\la\Big(4\nu\,\partial{\bf Q}_{mk}^v/\partial x_n
+2{\bf V}\times({\bf e}_n\times{\bf Q}_{mk}^v)
-2{\bf H}\times({\bf e}_n\times{\bf Q}_{mk}^h)\\
&+\,{\bf S}_m^v\times(\nabla_{\bf x}\times{\bf G}_{nk}^v
+{\bf e}_n\times{\bf S}_k^v)-{\bf S}_m^h\times(\nabla_{\bf x}
\times{\bf G}_{nk}^h+{\bf e}_n\times{\bf S}_k^h)\\
&+\,{\bf G}_{nk}^v\times(\nabla_{\bf x}\times{\bf S}_m^v)
-{\bf G}_{nk}^h\times(\nabla_{\bf x}\times{\bf S}_m^h)\Big)\cdot{\bf S}_j^{*v}
-2Q_{mk}^pS_{j,n}^{*v}\\
&+\Big(4\eta\,\partial{\bf Q}_{mk}^h/\partial x_n
+{\bf e}_n\times(2{\bf V}\times{\bf Q}_{mk}^h-2{\bf H}\times{\bf Q}_{mk}^v
+{\bf S}_m^v\times{\bf S}_k^h+{\bf S}_k^v\times{\bf S}_m^h)\\
&+\nabla_{\bf x}\times({\bf S}_m^v\times{\bf G}_{nk}^h
-{\bf S}_m^h\times{\bf G}_{nk}^v)\Big)\cdot{\bf S}_j^{*h}
+\Big(4\kappa\,\partial Q_{mk}^\theta/\partial x_n-2V_nQ_{mk}^\theta\\
&-({\bf G}_{nk}^v\cdot\nabla_{\bf x})S_m^\theta
-({\bf S}_m^v\cdot\nabla_{\bf x})G_{nk}^\theta
-S_{m,n}^vS_k^\theta\Big)S_j^{*\theta}
-\l(\nabla\widehat Q_{mkn}^v,\nabla\widehat Q_{mkn}^h,0)\cdot{\bf S}_j^*\Big\ra,\displaybreak\\
f_{nmkj}^3=&\Big\la\Big({\bf S}_n^v\times(\nabla_{\bf x}\times{\bf Q}_{mk}^v)
-{\bf S}_n^h\times(\nabla_{\bf x}\times{\bf Q}_{mk}^h)
+{\bf Q}_{mk}^v\times(\nabla_{\bf x}\times{\bf S}_n^v)\\
&-{\bf Q}_{mk}^h\times(\nabla_{\bf x}\times{\bf S}_n^h)\Big)\cdot{\bf S}_j^{*v}
+\Big(\nabla_{\bf x}\times({\bf S}_n^v\times{\bf Q}_{mk}^h
-{\bf S}_n^h\times{\bf Q}_{mk}^v)\Big)\cdot{\bf S}_j^{*h}\\
&-\Big(({\bf Q}_{mk}^v\cdot\nabla_{\bf x})S_n^\theta
+({\bf S}_n^v\cdot\nabla_{\bf x})Q_{mk}^\theta\Big)S_j^{*\theta}\Big\ra,
\end{align*}
$\delta_m^n$ is the Kronecker symbol, and
$\widehat G_{mkn}^v,~\widehat G_{mkn}^h,~\widehat Q_{mkn}^v$ and
$\widehat Q_{mkn}^h$ are space-periodic solutions to the Poisson equations
$$\nabla^2\widehat G_{mkn}^v=G_{mk,n}^v,\quad
\nabla^2\widehat G_{mkn}^h=G_{mk,n}^h,\quad
\nabla^2\widehat Q_{mkn}^v=Q_{mk,n}^v,\quad
\nabla^2\widehat Q_{mkn}^h=Q_{mk,n}^h.$$

The following identities for the coefficients $f$ prove useful:\\
1. For $n=3$ or 4, and any $m,k$ and $j$, straightforward algebra with
application of \rf{s34} and \rf{qmk} yields two relations:
$$f_{nmkj}^3+f_{nkmj}^3=2\la(\partial\l{\bf Q}_{mk}/\partial x_{n-2}-\l
(\partial{\bf Q}_{mk}/\partial x_{n-2}))\cdot{\bf S}_j^*\ra$$
and
$$f_{mnkj}^3+f_{mknj}^3+f_{kmnj}^3+f_{knmj}^3
=-2\la(\partial\l{\bf Q}_{mk}/\partial x_{n-2})\cdot{\bf S}_j^*\ra.$$
Consequently, no terms involving factors $c_{03}$ or $c_{04}$ are present
in the last sum in the r.h.s. of~\rf{rhs}.\\
2. For any $n,j=1,2$,
\BE f_{n34j}^2=f_{n43j}^2,\EE{f2}
since by inspection
$$f_{n34j}^2-f_{n43j}^2=\left\la\l
\left({\partial{\bf G}_{n3}\over\partial x_2}
-{\partial{\bf G}_{n4}\over\partial x_1}\right)\cdot{\bf S}_j^*\right\ra=0.$$

Derivation of equations for amplitudes is completed by performing
the following operations: ($i$) substitute in the r.h.s.~of \rf{nonev} and
\rf{rhs} the variables $c_{01}$ and $c_{02}$, using the second
relation in \rf{v0h0}; ($ii$) in the l.h.s.~of \rf{fslx}, eliminate the term,
proportional to $\partial C/\partial T$, by subtracting the appropriate
multiple of \rf{hsc}; ($iii$) introduce three new variables:
$\xi_1=C$, $\xi_2=q_1c_{03}+q_2c_{04}$ (the l.h.s.~of the modified \rf{fslx}
is proportional to $\partial\xi_2/\partial T$), and $\xi_3$ denoting the linear
combination of $C$, $c_{03}$ and $c_{04}$, whose derivative in $Y$ is
the l.h.s.~of \rf{nonev}; ($iv$) in the modified \rf{fslx}, replace
$\partial^2\xi_3/\partial Y^2$ by a derivative of a quadratic form of $\xi_i$
applying the non-evolutionary equation \rf{nonev}. Upon expressing
the r.h.s.~of \rf{hsc}, the modified \rf{fslx} and \rf{nonev} in terms
of $\xi_i$, we obtain with the use of \rf{f2} three equations,
\begin{align}
{\partial\xi_1\over\partial T}=&{\partial^2\over\partial Y^2}
\sum_{k=1}^3\beta_{1k}\,\xi_k+\gamma_1{\partial\xi_1^2\over\partial Y},\label{xi1}\\
{\partial\xi_2\over\partial T}=&{\partial^2\over\partial Y^2}(\beta_{21}\,\xi_1
+\beta_{22}\,\xi_2)+\sum_{i,j}\gamma_{2ij}\,\xi_i{\partial\xi_j\over\partial Y}
+\sigma\xi_1^3,\label{xi2}\\
{\partial\xi_3\over\partial Y}=&\sum_{i,j}\gamma_{3ij}\,\xi_i\xi_j
-\LLA\sum_{i,j}\gamma_{3ij}\,\xi_i\xi_j\RRA.
\label{xi3}\end{align}
The coefficients $\beta,\,\gamma$ and $\sigma$ can be calculated
following transformations ($i$)--($iv$) and using the expressions that we have
obtained for the coefficients of equations \rf{hsc}, \rf{fslx} and \rf{nonev}
in terms of solutions to the three auxiliary problems and eigenmodes from
the generalised kernel of $\l^*$, the adjoint to the operator of linearisation.

We find from \rf{xi1} that $\LA\xi_1\RA$ does not depend on the slow time;
hence, in view of \rf{means} for $n=0$ and the first equation in \rf{v0h0},
$\LA\xi_1\RA=0$. Relation \rf{xi3} does not completely specify $\xi_3$, since
it reduces to a trivial identity for the zero wave-number Fourier harmonic
(in $Y$) of \rf{xi3}. We thus lack an equation to close the system
\rf{xi1}--\rf{xi3}.

\subsubsection{The closing evolutionary equation}\label{clo}

To obtain the closing relation for the leading-order terms of the perturbation,
we have to consider the equations obtained at even higher orders. Such
a situation is often encountered in the dynamo theory, for instance,
one has to descend by 2 orders of magnitude to derive the magnetic
$\alpha$-effect in the classical almost-axisymmetric Braginsky dynamo \cite{Br}.

In order to derive the missing equation, we
average \rf{Oh} in~$Y$, apply condition \rf{means} for $n=2$, and find
\BE\sum_{k=3}^4{\partial\LA c_{0k}\RA\over\partial T}
\la{\bf S}_k\cdot{\bf S}_j^*\ra=\la\LA{\bf F}\RA\cdot{\bf S}_j^*\ra.\EE{biav}
The component of \rf{Oh}, parallel to $\bf q$, has been fully explored
in Section~\ref{add}; we consider now the component, perpendicular
to $\bf q$. Scalar multiplication of \rf{biav} by ${\bf q}^\perp$
and the use of \rf{SkSj} yields
\BE\sum_{j=1}^2\left(\sum_{k=1}^2u'_{kj}{\partial\over\partial T}\LA c_{0,k+2}\RA
+\la\LA{\bf F}\RA\cdot{\bf S}_j^*\ra\right)(-1)^jq_{3-j}=0.\EE{mcl}
We express here the mean of \rf{rhs},
$$\la\LA{\bf F}\RA\cdot{\bf S}_j^*\ra=\sum_{n=1}^2\sum_{m=1}^4\sum_{k=1}^4
\LLA{\partial c_{0k}\over\partial X_n}c_{0m}\RRA f_{nmkj}^2
+\sum_{n=1}^2\sum_{m=1}^2\sum_{k=1}^2\LA c_{0n}c_{0m}c_{0k}\RA f_{nmkj}^3,$$
in terms of $\xi_i$ using \rf{f2}:
$${\partial\over\partial T}\,\LA\mu'_1\xi_2+\mu'_2\xi_3\RA=
\LLA\xi_1{\partial\over\partial Y}(\gamma'_2\xi_2+\gamma'_3\xi_3)\RRA
+\sigma'\LA\xi_1^3\RA$$
and subtract from this equation \rf{xi2} averaged in $Y$ and multiplied
by $\sigma'/\sigma$. This finally yields the closing equation:
\BE{\partial\over\partial T}\,\LA\mu_1\xi_2+\mu_2\xi_3\RA=
\LLA\xi_1{\partial\over\partial Y}(\gamma_{42}\xi_2+\gamma_{43}\xi_3)\RRA.\EE{xi4}
Equations \rf{xi1}--\rf{xi3} and \rf{xi4} comprise a closed system in
$\xi_i$, equivalent to the system of amplitude equations.

\section{Numerical results}\label{resu}

In this section, we present results of numerical investigation
of the two-scale weakly nonlinear stability of space-periodic steady-state
convective dynamos in a horizontal layer of rotating conducting fluid,
that comprise the branch labelled S$^{\rm R1}_8$ in \cite{CGPZ}. The branch
is parameterised by the Taylor number, Ta, and exists in the interval
$673.7\le{\rm Ta}\le 704$. These convective MHD steady states are symmetric
about the vertical axis, $x_3$, and hence the significant (acting
on the amplitudes) $\alpha$-effect is absent. Dynamos from this branch have
the group of symmetries ${\bf D}_2$, the second group generator being
the composition of reflection about the midplane and the shift by half a period
in the horizontal direction $x_1$. For this group of symmetries, the matrix
$\U$ is non-zero, this invalidating the multiscale analysis \cite{vz}
of large-scale perturbations of these dynamos.

We have refined the steady states from the branch S$^{\rm R1}_8$
with the resolution of $64^2\times32$ Fourier harmonics. When solving
the auxiliary problems and computing neutral eigenmodes
of $\l^*$ (the adjoint to the operator of linearisation), solutions
in the form of truncated Fourier series have been sought.
Pseudospectral methods \cite{go} have been used (resolution: $96^2\times 48$
Fourier harmonics before dealiasing) for computing various products
encountered in the operator~$\l$. The code \cite{Fok} realising
the biconjugate gradients stabilised method BiCGstab($\ell$)
\cite{SF,SV95,SV96} for $\ell=2$ has been applied to solve the linear system
of equations for the Fourier coefficients. Table~\ref{tab0} shows the values
of the coefficients in the system of amplitude equations for ${\rm Ta}=675$.

\begin{table}[b!]
\caption{Coefficients in the amplitude equations for ${\rm Ta}=675$.}\label{tab0}
\center\begin{tabular}{|c|c|l|}\hline
Eq.&Coeff.&Values\\\hline
\rf{xi1}&$\beta_{1k}$&
0.884249123107103,~~0.973164220817207,~~0.002542972607168343\\
&$\gamma_1$& 0.588162048986537\\\hline
\rf{xi2}&$\beta_{2k}$&0.366161979311609,~~4.15859432509866\\
&$\gamma_{2ij}$&
0.806971618405987,~~-4.23666924855939,~~0.158691136470146,\\
&&-4.82055714849942,~~-42.2469280336739,~~0.135854143000776,\\
&&0.209741278940542,~~0.135854143000747,~~0.003874598372548109,\\
&$\sigma$&0.192034252840662\\\hline
\rf{xi3}&$\gamma_{3ij}$&
-7.08024457707350,~~8.63396618933175,~~-0.500706409449314,\\
&&8.63396618933175,~~85.1114936402045,~~0.385556218828258,\\
&&-0.500706409449314,~~0.385556218828260,~~-0.03382831817117658\\\hline
\rf{xi4}&$\mu_i$&1,~~-0.002659134450504667\\
&$\gamma_{4i}$&0.596884895067222,~~-0.04846657396378542\\\hline
\end{tabular}\end{table}

\begin{table}[b!]
\caption{Eigenvalues of matrix \rf{para}, $d_i$, for two unit vectors $\bf q$
(such that ${\bf q}^\perp$ are eigenvectors of matrix $\U$).}\label{tab1}
\center\begin{tabular}{|c|c|c|c|}\hline
Ta&$\bf q$&$d_1$&$d_2$\\\hline
673.7&${\bf q}^1=(0.998515,~0.054469)$&0.763328&4.298197  \\
     &${\bf q}^2=(0.822115,~0.569320)$&0.264885&-261.047947\\\hline
675  &${\bf q}^1=(0.998613,~0.052645)$&0.778817&4.264026  \\
     &${\bf q}^2=(0.825155,~0.564905)$&0.270588&-361.244316\\\hline
680  &${\bf q}^1=(0.998954,~0.045718)$&0.889724&4.175111  \\
     &${\bf q}^2=(0.836586,~0.547835)$&0.289251&1326.420209\\\hline
690  &${\bf q}^1=(0.999477,~0.032314)$&1.588269&4.598703  \\
     &${\bf q}^2=(0.858485,~0.512837)$&0.297058&201.124913\\\hline
700  &${\bf q}^1=(0.999808,~0.019574)$&2.518614&16.785099 \\
     &${\bf q}^2=(0.879679,~0.475566)$&0.174285&212.326992\\\hline
703  &${\bf q}^1=(0.999873,~0.015897)$&2.597766&65.565570 \\
     &${\bf q}^2=(0.886016,~0.463653)$&0.057593&281.662699\\\hline
\end{tabular}\end{table}

\subsection{Linear stability of constant amplitudes}\label{coam}

The behaviour of solutions depends crucially on whether the evolutionary
equations \rf{xi1}--\rf{xi2} are parabolic. They constitute a parabolic
(sub)system, when real parts of both eigenvalues of the matrix
\BE\displaystyle\left[\begin{matrix}\beta_{11}&
\!\!\beta_{12}\\\beta_{21}&\!\!\beta_{22}\end{matrix}\right]\EE{para}
are positive.
Otherwise, the second-order partial differential operator in the r.h.s.~of
\rf{xi1}--\rf{xi2} would cause a superexponential growth of solutions
in time (unless it is suppressed by nonlinear terms --- but
they are more likely to boost rather than inhibit this growth).
The term $\beta_{13}\partial^2\xi_3/\partial Y^2$ is not involved in this
analysis, because \rf{xi3} is a non-evolutionary equation. The eigenvalues
$d_i$ shown in Table~\ref{tab1} reveal a high variability
of the parameter values controlling the perturbation despite
the interval, in Ta, of existence of the branch is short. At the left end
of this interval for S$^{\rm R1}_8$, $d_1>0$ but $d_2<0$,
indicating a rapid (in the slow time) instability. In the remaining part
of the interval, both $d_i>0$
and hence the second-order operator acts as the stabilising diffusion.

\begin{figure}[t]
\center\includegraphics[width=.666\textwidth]{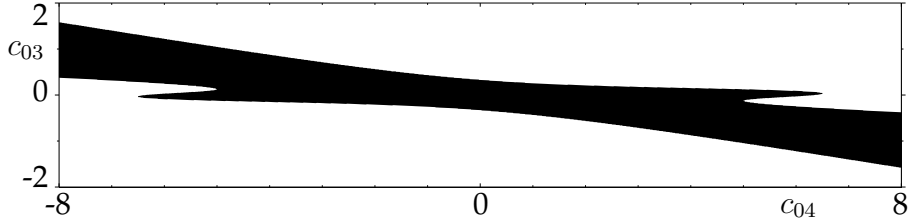}

\vspace*{-.15\textwidth}\hspace*{-.66\textwidth}$c_{03}$

\vspace*{.09\textwidth}\hspace*{.5\textwidth}$c_{04}$

\caption{The region of linear stability (black) of constant solutions
to the system of amplitude equations in the rectangle
$|c_{03}|\le2,~|c_{04}|\le8$,~$C=0$.}
\label{sta}\end{figure}

Clearly, $\xi_1=0$ and any constant in time and space $\xi_2$ and $\xi_3$
(or, equivalently, $C=0$ and any constant $c_{03}$ and $c_{04}$) are
solutions to the system of amplitude equations \rf{xi1}--\rf{xi3}, \rf{xi4}.
In other words, any linear combination of the neutral modes ${\bf S}_{k+2}$
\rf{s34} with constant coefficients is a solution to the stability
problem in the broader class of weakly nonlinear large-scale perturbations.
The condition of parabolicity of the linear operator in \rf{xi1}--\rf{xi2} is
the condition of stability of the solution $\xi_i=0$ (i.e., $C=c_{03}=c_{04}=0$).
A constant solution $(0,\xi_2,\xi_3)$ is linearly stable
if ${\rm Re}\varrho<0$ for both roots $\varrho$ of the quadratic equation
$$\det\left[\begin{matrix}
\beta_{11}+\varrho&\beta_{12}&\beta_{13}\\
\i\beta_{21}+\Gamma_{21}&\i\beta_{22}+\Gamma_{22}+\i\varrho&\Gamma_{23}\\
\Gamma_{31}&\Gamma_{32}&-1\\\end{matrix}\right]=0$$
(see Fig.~\ref{sta}). Here we have denoted
$$\Gamma_{2k}=\sum_{j=2}^3\gamma_{2jk}\xi_j,\qquad
\Gamma_{3k}={\displaystyle\left(\sum_{j=2}^3(\gamma_{3jk}+\gamma_{3kj})\xi_j\right)
\left(\sum_{j=2}^3(\gamma_{3j3}+\gamma_{33j})\xi_j\right)
-\i\sum_{j=2}^3(\gamma_{3jk}+\gamma_{3kj})\xi_j\over\displaystyle
1+\left(\sum_{j=2}^3(\gamma_{3j3}+\gamma_{33j})\xi_j\right)^{\!\!2}}.$$

\subsection{Two types of behaviour of amplitudes}\label{beam}

We report now results of numerical investigation of the system of amplitude
equations \rf{xi1}--\rf{xi3}, \rf{xi4}. Solutions,
$2\pi$-periodic in $Y$, have
been sought as truncated Fourier series in the spatial variable.
Pseudospectral methods \cite{go} have been applied. The presence
of the cubic term in \rf{xi2} necessitates performing the dealiasing
by discarding a half of the highest-wave-number harmonics, rather than 1/3
of them as stipulated by the standard 3/2 rule \cite{or} (when evaluating
quadratic terms). The second-order implicit midpoint method has been used
to integrate the system in time. In terms of the Fourier coefficients
of the variables $\xi_s(Y,T)$ denoted by
$\widehat{\bxi}(T)=(\widehat{\xi}_1(T),...,\widehat{\xi}_{3M}(T))$, the system
takes the form
\begin{align}
{\partial\widehat{\xi}_m\over\partial T}=f_m(\widehat{\bxi}),\qquad&
1\le m\le2M+1,\label{no1}\\
0=f_m(\widehat{\bxi}),\qquad&2M+2\le m\le3M\label{no2}
\end{align}
(here evolutionary equations \rf{no1} stem from \rf{xi1}, \rf{xi2} and \rf{xi4},
and \rf{no2} from \rf{xi3}). In this notation, the midpoint method propagates
solution $\widehat{\bxi}^{(n)}$ at time $t_n$ a step forward to
$\widehat{\bxi}^{(n+1)}$ at time $t_{n+1}=t_n+dt$ according to the equations
\begin{align}
\widehat{\xi}_m^{(n+1)}-\widehat{\xi}_m^{(n)}-dt\,f_m
\left({\widehat{\bxi}^{(n+1)}+\widehat{\bxi}^{(n)}\over2}\right)=0,\qquad&
1\le m\le2M+1,\label{u1}\\
f_m(\widehat{\bxi}^{(n+1)})=0,\qquad&2M+2\le m\le3M.\label{u2}
\end{align}
To solve \rf{u1}--\rf{u2}, we have used the routine {\tt newt} \cite{NR}
that realises a quasi-Newton method (which is a combination of the Newton
method with a minimisation technique). The initial conditions are required
to satisfy \rf{u2}. Mostly, a time step $dt=10^{-4}$ was used.
A number of runs have been also performed using the standard
fourth-order Runge--Kutta method and the time step $dt=10^{-3}$, with
\rf{u2} being solved at each substep. Comparison of the results obtained
by the two methods has shown a satisfactory quantitative agreement.

Solutions to amplitude equations \rf{xi1}--\rf{xi3} and \rf{xi4} apparently
do not obey any conservation law and thus, in principle, can saturate
at constant values or grow to infinity. In fact, we have not found
any other patterns of behaviour.

\setlength{\unitlength}{.55mm}
\begin{figure}[p]
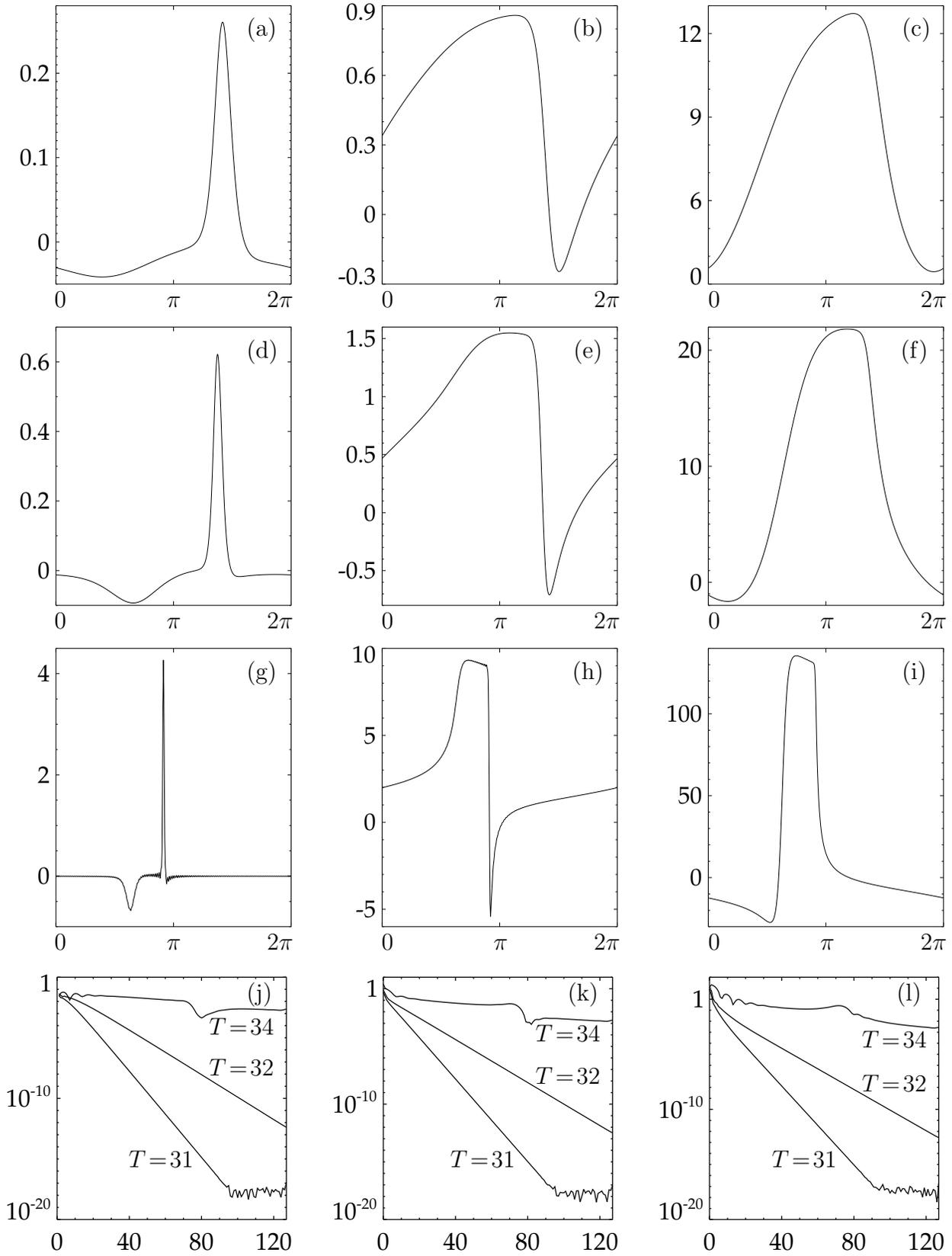

\begin{picture}(300,400)(0,0)
\put(0,310){\hfill\includegraphics[width=5cm]{prof31_c}~~~~
\includegraphics[width=5cm]{prof31_c03}~~~~
\includegraphics[width=5cm]{prof31_c04}\hfill}
\put(0,304){\hspace*{8.5mm}0\hspace*{17.5mm}$\pi$\hspace*{15mm}$2\pi$~~~~
\hspace*{8.5mm}0\hspace*{17.5mm}$\pi$\hspace*{15mm}$2\pi$~~~~
\hspace*{8.5mm}0\hspace*{17.5mm}$\pi$\hspace*{15mm}$2\pi$}
\put(0,390){\hspace*{42mm}(a)\hspace*{51.5mm}(b)\hspace*{52mm}(c)}

\put(0,208){\hfill\includegraphics[width=5cm]{prof33_c}~~~~
\includegraphics[width=5cm]{prof33_c03}~~~~
\includegraphics[width=5cm]{prof33_c04}\hfill}
\put(0,202){\hspace*{8.5mm}0\hspace*{17.5mm}$\pi$\hspace*{15mm}$2\pi$~~~~
\hspace*{8.5mm}0\hspace*{17.5mm}$\pi$\hspace*{15mm}$2\pi$~~~~
\hspace*{8.5mm}0\hspace*{17.5mm}$\pi$\hspace*{15mm}$2\pi$}
\put(0,288){\hspace*{42mm}(d)\hspace*{51.5mm}(e)\hspace*{52mm}(f)}

\put(0,106){\hfill\includegraphics[width=5cm]{prof34_c}~~~~
\includegraphics[width=5cm]{prof34_c03}~~~~
\includegraphics[width=5cm]{prof34_c04}\hfill}
\put(0,100){\hspace*{8.5mm}0\hspace*{17.5mm}$\pi$\hspace*{15mm}$2\pi$~~~~
\hspace*{8.5mm}0\hspace*{17.5mm}$\pi$\hspace*{15mm}$2\pi$~~~~
\hspace*{8.5mm}0\hspace*{17.5mm}$\pi$\hspace*{15mm}$2\pi$}
\put(0,186){\hspace*{42mm}(g)\hspace*{51.5mm}(h)\hspace*{52mm}(i)}

\put(0,4){\hfill\includegraphics[width=5cm]{spec_c}~~~~
\includegraphics[width=5cm]{spec_c03}~~~~
\includegraphics[width=5cm]{spec_c04}\hfill}
\put(0,84){\hspace*{42mm}(j)\hspace*{51.5mm}(k)\hspace*{52mm}(l)}
\put(64.5,73){$T\!=\!34$}
\put(168,71){$T\!=\!34$}
\put(271.5,69){$T\!=\!34$}
\put(64.5,60){$T\!=\!32$}
\put(168,57.5){$T\!=\!32$}
\put(271.5,55){$T\!=\!32$}
\put(39,31){$T\!=\!31$}
\put(141,31){$T\!=\!31$}
\put(243,31){$T\!=\!31$}
\end{picture}
\caption{The amplitudes (vertical axis) $C$ (panels (a), (d), (g)),
$c_{03}$ ((b), (e), (h)) and $c_{04}$ ((c), (f), (i)) at $T=31$ ((a)--(c)),
$T=33$ ((d)--(f)) and $T=34$ ((g)--(i)); horizontal axis: $Y$. Absolute values
of the Fourier coefficients of the amplitudes (vertical axis) $C$ (j), $c_{03}$
(k) and $c_{04}$ (l) at the same times $T$; horizontal axis: wave number.}
\label{zub}\end{figure}

Let us now discuss in detail a run for ${\rm Ta}=675$ and ${\bf q}={\bf q}^1$
(see Table~\ref{tab1}), which we will refer to as ``the main sample run''.
In a preparation of a blow-up, maxima of all the three amplitudes increase
in time indefinitely; a graph of an amplitude near its maximum turns into
a spike, travelling along the $Y$-axis at an increasing speed and finally
growing to very large values at a finite time (see a typical behaviour
in Fig.~\ref{zub} and a video stored at arXiv.org as an ancillary file
Ta675.mp4). Simulated solutions
were truncated to 256 Fourier harmonics in the spatial variable (i.e.,
they involved wave numbers up to 127, and computation
of nonlinear terms involved 512 harmonics before dealiasing). The noise
in the graphs of absolute values of the Fourier coefficients of the amplitudes
(panels (j)--(l) in Fig.~\ref{zub}) at $T=31$ for wave numbers exceeding 95
indicates, that such a resolution is excessive when computing with the double
precision up to this time. In the course of evolution, the energy gets
eventually almost evenly distributed between all Fourier harmonics (panels
(j)--(l) in Fig.~\ref{zub}); at the final stage, amplitude $C$ resembles
a periodically-replicated $\delta$-function moving at a constant speed.
(Of course, at this stage the numerical solution is underresolved and
cannot be fully trusted any more.) A common
feature of nonlinear partial differential equations of the first order such as
the inviscid Burgers equation is development of a jump due to intersection
of characteristics; in the present case the singularity can be rather
interpreted as a derivative of a function, experiencing a jump.
However, addition of a small diffusivity into the Burgers
equation smooths out the jumps, which is not the case in the present system
of amplitude equations, where diffusion is present and is not infinitesimal.

\begin{figure}[p]
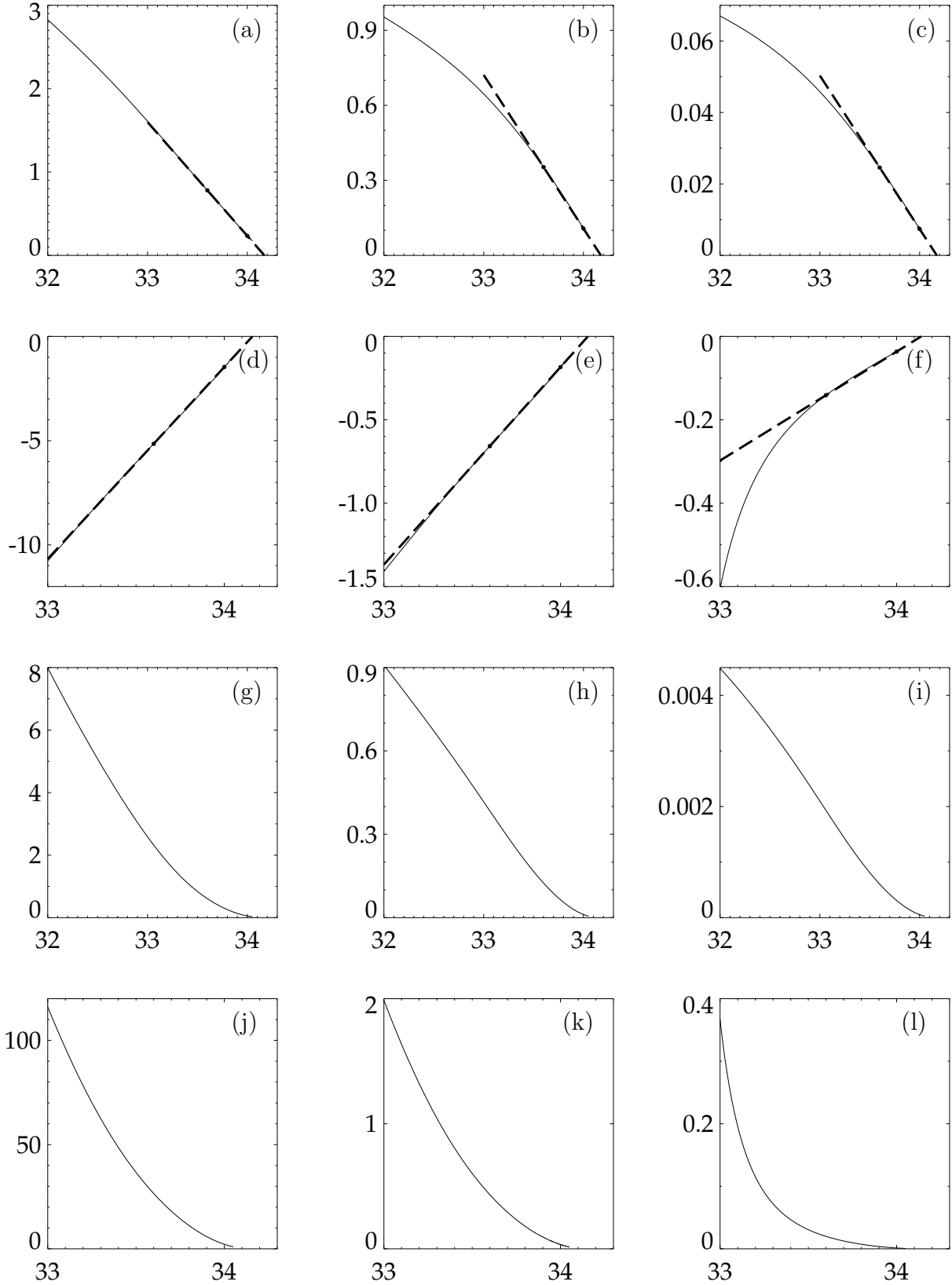

\begin{picture}(300,400)(0,0)
\put(0,310){\hfill\includegraphics[width=5cm]{max1_c}~~~~
\includegraphics[width=5cm]{max1_c03}~~~~
\includegraphics[width=5cm]{max1_c04}\hfill}
\put(0,390){\hspace*{42mm}(a)\hspace*{51.5mm}(b)\hspace*{52mm}(c)}

\put(0,208){\hfill\includegraphics[width=5cm]{min1_c}~~~~
\includegraphics[width=5cm]{min1_c03}~~~~
\includegraphics[width=5cm]{min1_c04}\hfill}
\put(0,288){\hspace*{43mm}(d)\hspace*{51.5mm}(e)\hspace*{51mm}(f)}

\put(0,106){\hfill\includegraphics[width=5cm]{max2_c}~~~~
\includegraphics[width=5cm]{max2_c03}~~~~
\includegraphics[width=5cm]{max2_c04}\hfill}
\put(0,186){\hspace*{42mm}(g)\hspace*{51.5mm}(h)\hspace*{52mm}(i)}

\put(0,4){\hfill\includegraphics[width=5cm]{min2_c}~~~~
\includegraphics[width=5cm]{min2_c03}~~~~
\includegraphics[width=5cm]{min2_c04}\hfill}
\put(0,84){\hspace*{42mm}(j)\hspace*{51.5mm}(k)\hspace*{52mm}(l)}
\end{picture}
\caption{The quantities \rf{mami} (vertical axis) $M_{f,\max}(T)$ (panels
(a)--(c)), $M_{f,\min}(T)$ ((d)--(f)), $M^2_{f,\max}(T)$ ((g)--(i)) and
$M^2_{f,\min}(T)$ ((j)--(l)) for $C$ ((a), (d), (g), (j)), $c_{03}$
((b), (e), (h), (k)) and $c_{04}$ ((c), (f), (i), (l)) during the preparation
of the blow-up of a solution to the system of amplitude equations. Horizontal
axis: the slow time $T$. Dashed lines on panels (a)-(f): lines through
the solid points on the respective graphs, approximating the asymptotes
that characterise the final stage of the process of preparation.}
\label{mxn}\end{figure}

Blow-ups are well-known to be linked with self-similar
solutions (see, e.g., \cite{Bar,egg}). It is easy to check that equations
\rf{xi1}--\rf{xi3} and \rf{xi4} admit self-similar solutions of the form
\BE(C,c_{03},c_{04})=(T_\star-T)^{-1/2}\,\Phi(Y(T_\star-T)^{-1/2})\EE{ass}
(see \cite{pol}). To examine, whether the solution has this asymptotics near
the singularity, we plot in Fig.~\ref{mxn} graphs of the quantities
\BE M_{f,\max}(T)=1/\max_Yf(Y,T),\qquad M_{f,\min}(T)=1/\min_Yf(Y,T),\EE{mami}
as well $M^2_{f,\max}(T)$ and $M^2_{f,\min}(T)$; here the function $f$ is any
amplitude $C$, $c_{03}$ and $c_{04}$. (The maxima and their positions are
determined applying the quadratic interpolation.) The graphs reveal that
the factor $(T_\star-T)^{-1/2}$ may indeed correctly describe the growth of
the amplitudes at intermediate times preceding the blow-up
($31\le T\le33$ for the maxima of the solution under consideration, as seen
in panels (g)--(i) of Fig.~\ref{mxn}). (This is not seen as clearly
in panels (j)--(l) of Fig.~\ref{mxn} for the minima, because
$\min_Yc_{04}(Y,T)$ vanishes at $T\approx33.66$, forcing us
to plot $M_{f,\min}(T)=1/\min_Yf(Y,T)$ for a shorter time interval.) However,
further on, at times immediately preceding the blow-up, the maxima and minima
behave as $(T_\star-T)^{-1}$ (as follows from graphs in panels (a)--(f)
of Fig.~\ref{mxn}). The graphs in panels (a)--(c) for the maxima of amplitudes
and in panels (d)--(f) for their minima coherently indicate that
the blow-up occurs at $T_\star\approx34.17$\,.

\begin{figure}[t]
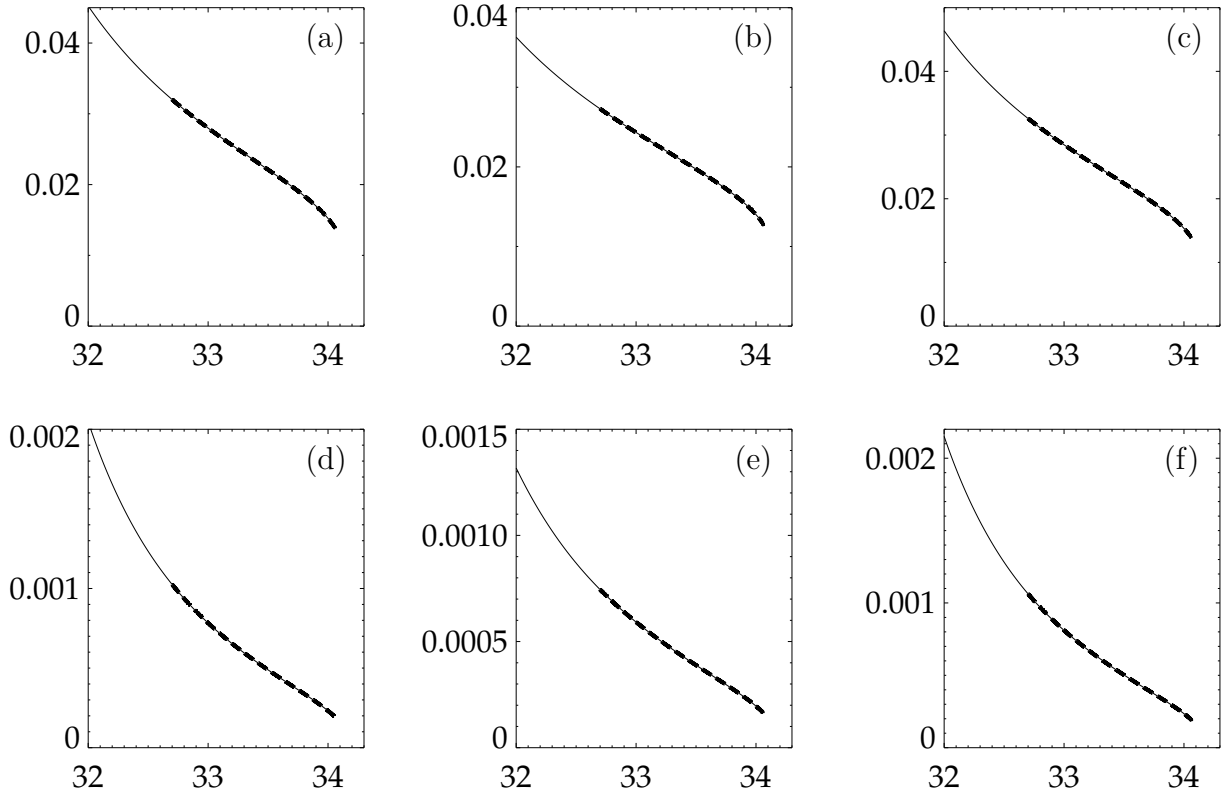

\begin{picture}(300,200)(0,0)
\put(0,106){\hfill\includegraphics[width=5cm]{max1co_c}~~~~
\includegraphics[width=5cm]{max1co_c03}~~~~
\includegraphics[width=5cm]{max1co_c04}\hfill}
\put(0,186){\hspace*{42mm}(a)\hspace*{51.5mm}(b)\hspace*{52mm}(c)}

\put(0,4){\hfill\includegraphics[width=5cm]{max2co_c}~~~~
\includegraphics[width=5cm]{max2co_c03}~~~~
\includegraphics[width=5cm]{max2co_c04}\hfill}
\put(0,84){\hspace*{42mm}(d)\hspace*{51.5mm}(e)\hspace*{52mm}(f)}
\end{picture}
\caption{Positions of the maxima (vertical axis), $Q_f(T)$, (panels (a)--(c))
and $Q^2_f(T)$ ((d)--(f)) for $C$ ((a), (d)), $c_{03}$ ((b), (e)) and $c_{04}$
((c), (f)) during the preparation of the blow-up of a solution to the system
of amplitude equations. Horizontal axis: the slow time $T$. Dashed lines: same
quantities computed using a double spatial resolution and a 10 times smaller
time step.}\label{mpo}\end{figure}

\begin{figure}[t]
\center\includegraphics[width=.666\textwidth]{mean}

\vspace*{-.255\textwidth}\hspace*{-.666\textwidth}$\LA c_{03}\RA$

\vspace*{.205\textwidth}\hspace*{.44\textwidth}$\LA c_{04}\RA$

\caption{Isolines of blow-up times for varying initial $\LA c_{03}\RA$ and
$\LA c_{04}\RA$. The region, for which solutions tend to constant values,
is cross-hatched. When a point $(\LA c_{03}\RA,\LA c_{04}\RA)$ approaches
this region, the times of blow-up tend to infinity, and hence there are
isolines for the values 10 and higher near the upper border of this
region, and for values 1 and higher near the lower border. These isolines are
very close to the respective borders and are not shown.
The region of linear stability of constant solutions
to the system of amplitude equations is shown in gray (cf.~Fig.~\ref{sta}).}
\label{mea}\end{figure}

To see how well the self-similar solution \rf{ass} approximates
near the singularity the spatial dependence of the amplitudes, we have explored
the behaviour of $P_f(T)$, the position of the maximum of $f(Y,T)$ at time $T$,
where as function $f$ we again take the amplitudes $C$, $c_{03}$
and $c_{04}$. For the self-similar ansatz \rf{ass} we would obtain
\BE P_f(T)=P_\Phi\,(T_\star-T)^{1/2},\EE{maxp}
where $P_\Phi$ is the argument at which the profile $\Phi$ in \rf{ass} takes
the maximum value; we therefore expect $P^2_f(T)$ to be a linear function of
time. Graphs of positions of the maxima, $P_f(T)$, as well as $P^2_f(T)$
as functions of the slow time $T$ are plotted in Fig.~\ref{mpo}, and they do not
contradict the hypothesis, that asymptotically \rf{maxp} holds in the time
interval $33\le T\le34$. However, closer to the time of singularity $T_\star$,
the graphs of $P^2_f(T)$ (panels (d)--(f) of Fig.~\ref{mpo}) bend down.
To check that this bending is genuine and not caused
by the lack of resolution, we have recomputed the solution with a twice higher
spatial resolution and a 10 times smaller time step; the graphs obtained
for the refined positions are shown in Fig.~\ref{mpo} by dashed lines, which
visually coincide with the original graphs. Note, that the graphs
of $P^2_f(T)$ (panels (a)--(c) of Fig.~\ref{mpo}) feature a similar behaviour,
which is close to a linear one at a larger time interval; however, $P_f(T)$
bends down near $T_\star$ more visibly. Comparison of panels (a)--(c) versus
panels (d)--(f) shows, that it is difficult to make a reliable conclusion
about the exponent in the argument of $\Phi$ in \rf{maxp} based on this data.

We have investigated how the occurrence and the time of blow-up depend on
the mean amplitudes of the initial conditions (see Fig.~\ref{mea}). The mean
amplitudes $\LA c_{03}\RA$ and $\LA c_{04}\RA$ of the two neutral zero-mean
small-scale stability modes, that exist due to translation invariance
in horizontal directions of equations of convective dynamo, appear to be
important control parameters: if in the course of temporal evolution these
means quit the region of linear stability of constant solutions (see
the previous subsection), typically, the solution eventually experiences
a blow-up at a finite time.

Ideally, in such an investigation we would like to compare solutions
to the system \rf{xi1}--\rf{xi3} and \rf{xi4} obtained for initial conditions
where these means are varied and the fluctuating parts are kept unaltered.
However, this is impossible, because the initial data must satisfy
the non-evolutionary equation \rf{xi3}. Consequently, Fig.~\ref{mea} is plotted
for solutions with the initial conditions constructed
as follows: At $T=0$, the fluctuating parts of the amplitudes $c_{03}(Y)$ and
$c_{04}(Y)$ are the same in all the runs (they coincide, respectively,
with the fluctuating parts of $c_{03}$ and $c_{04}$ in the discussed above main
sample run at $T=30$; the r.m.s.~means of the fluctuating parts are,
respectively, 0.22 and 1.88). The zero-mean amplitude $C(Y)$ is then found
from \rf{xi3} for each pair $\LA c_{03}\RA,\LA c_{04}\RA$ separately.

We see in Fig.~\ref{mea} that the region of mean amplitudes, for which
solutions tend to constant values, is within the region of linear
stability of constant solutions (see subsection~\ref{coam}). We also observe
that when a point $(\LA c_{03}\RA,\LA c_{04}\RA)$ approaches the lower boundary
of this region outside it, the times of blow-up tend to infinity much faster than
when the upper boundary is approached, suggesting that the blow-up preparation
proceeds differently, at least in details, in the two areas above and below
the region of linear stability of constant solutions. Simulations
support this conclusion: For the main sample run,
$\LA c_{03}\RA=0.45,\,\LA c_{04}\RA=8.02$ at $T=30$; it corresponds to a point
in the area above the region of stability of constant solutions in
Fig.~\ref{mea}. An additional run for ${\rm Ta}=675$ and the initial conditions,
that have been constructed as explained above for $\LA c_{03}\RA=-1.2$ and
$\LA c_{04}\RA=6$, is represented in Fig.~\ref{mea} by a point in the area
below the region (a video is stored at arXiv.org as an ancillary file
Ta675a.mp4). In both runs the graphs of the moduli of Fourier coefficients
of the amplitudes versus the wave number $k$
remain near the blow-up predominantly linear. However, while in the main
sample run their steepness monotonically decreases in time, in the additional
run the steepness experiences a complex oscillatory behaviour. Also,
while at the final stage one spike of a large height develops near the time
of singularity in the main sample run, high-amplitude spatial oscillations
burst out in space in the additional run.

\subsection{Dependence on the Taylor number}\label{dTa}

We summarise now the numerical results obtained for several other
steady states from the branch S$^{\rm R1}_8$. Although the coefficients
of the amplitude equations significantly vary within the branch (most of them
by more than an order of magnitude), we do not encounter any other
types of the temporal behaviour of amplitudes, rather than those found for
${\rm Ta}=675$: the development of a singularity at a finite time,
or the decay of coefficients to constant values, i.e., degeneration of
a large-scale perturbation to a neutral zero-mean short-scale mode
existing due to the spatial invariance of the convective system.

\begin{figure}[t!]
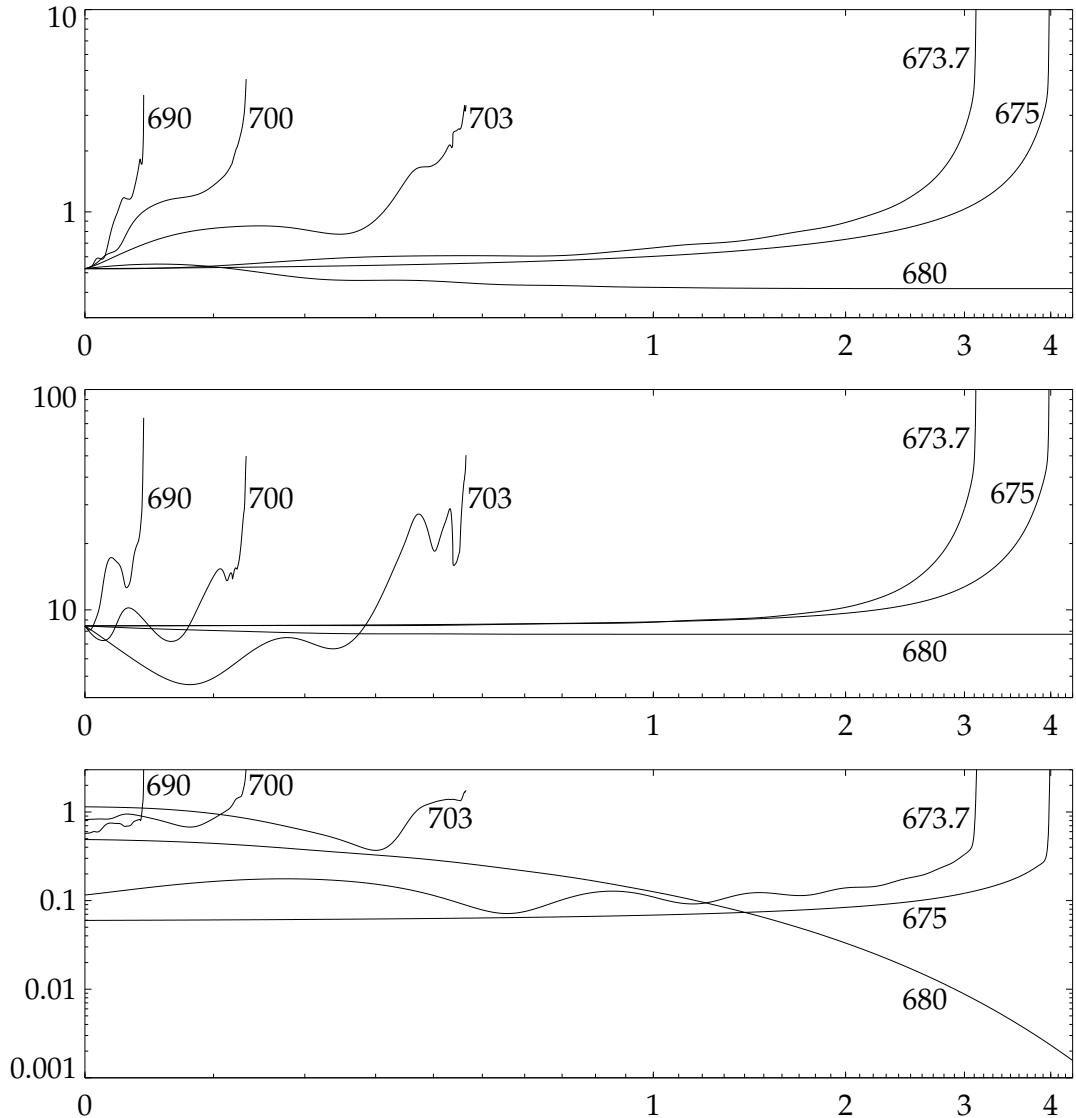

\center\includegraphics[width=.8\textwidth]{ec03}
\center\includegraphics[width=.8\textwidth]{ec04}
\center\includegraphics[width=.8\textwidth]{eC}
\caption{Lebesgue norms (vertical axis) of amplitudes $c_{03}$ (upper panel),
$c_{04}$ (middle panel) and $C$ (lower panel) as functions of the slow time
$T$ (horizontal axis; the scale is logarithmic to have a better
display of details of the graphs for ${\rm Ta}\ge690$, for which
the blow-up times are small) for ${\bf q}={\bf q}^1$ (see Table~\ref{tab1}).
The plots are labelled by the respective Taylor number values.}
\label{eq1}\end{figure}

The genericity of these two types of behaviour is illustrated by Fig.~\ref{eq1},
showing the temporal evolution of the Lebesgue norms of amplitudes
(defined as $\sqrt{\LA c^2(Y,T)\RA}$ for an amplitude $c(Y,T)$)
for several values of the Taylor number. The graphs in Fig.~\ref{eq1}
are plotted for the initial conditions constructed essentially by the same
procedure as in the previous subsection: At $T=0$, amplitudes $c_{03}(Y)$ and
$c_{04}(Y)$ are the same in the runs for all considered Ta, and coincide
with those in the main sample run at $T=30$. The zero-mean amplitude $C(Y)$ is
found from \rf{xi3} for each Taylor number separately.

Figure~\ref{eq1} shows that the blow-up times depend on the Taylor
number non-monotonically. Although for ${\rm Ta}\ge690$ the employed initial
profiles of $c_{03}$ and $c_{04}$ quickly lead to a blow-up,
for the nearby value ${\rm Ta}=680$ the singularity does not develop
and the perturbation tends to a short-scale zero-mean neutral mode.
Noteworthily, prior to the final stage amplitude $C$ remains close to unity,
and therefore the cubic nonlinearity of the system of amplitude
equations, which is associated with $\xi_1=C$ (see \rf{xi2}), apparently
does not play an important role in the development of the singularity.

Further qualitative information is revealed by the videos of the final stage
of the preparation of the blow-up in two runs for ${\rm Ta}=680$ and 703
(for initial conditions constructed by an alternative procedure),
and in the additional run for ${\rm Ta}=675$ (in all the three runs
${\bf q}={\bf q}^1$), which are stored at arXiv.org as ancillary files
Ta680.mp4, Ta703.mp4 and Ta675a.mp4, respectively.
For ${\rm Ta}=703$, like in the runs for ${\rm Ta}=675$,
the graphs of the moduli of Fourier coefficients of the amplitudes as
functions of the wave number $k$ remain predominantly linear, although the fine
structure of these graphs is much richer than in the case of the main sample
run --- all amplitudes have now several extrema, whose magnitudes grow in time.
In the main sample run for ${\rm Ta}=675$, the steepness
of such graphs just decreases in time; by contrast, for ${\rm Ta}=703$
initially the steepness increases, so that the round-off level $\sim10^{-15}$
is attained at $k\sim60$ (in this simulation, solutions have been truncated
to 512 Fourier harmonics in $Y$ after dealiasing), and afterwards the slope
starts to decrease; in the additional run for ${\rm Ta}=675$ the slope
experiences an oscillatory behaviour. Eventually the slope
of the graphs decreases until near the time of singularity the graphs
become close to horizontal ones. The growth of the extrema (in absolute values)
of the amplitudes intensifies in time, but none of the extrema gives rise
to a $\delta$-function, as the single maximum of $C$ does in the main
sample run for ${\rm Ta}=675$. For ${\rm Ta}=703$, just before the emergence
of the singularity, a secondary fast moving perturbation of amplitudes
develops, which modifies their so far preserved forms; the overall linear shape
of the graphs of absolute values of the Fourier coefficients becomes
deformed. In the additional run for ${\rm Ta}=675$,
high-frequency high-amplitude spatial oscillations are observed.
However, the spatial resolution becoming insufficient,
the occurrence of these final events cannot be fully trusted.

The evolution of perturbations near the singularity for ${\rm Ta}=680$
significantly differs from those for ${\rm Ta}=675$ and 703. The singularity
sets in much faster. Two and then three global maxima develop in graphs
of the moduli of Fourier coefficients of the amplitudes versus the wave number,
which are initially predominantly linear. Each amplitude has initially one
maximum; the maxima do not yield $\delta$-functions, but rather burst into
packets of amplitude-modulated blobs of high-frequency oscillations.

\section{Concluding remarks}\label{conclu}

We have performed a two-scale analysis of large-scale perturbations of
a periodic array of steady thermal convective hydromagnetic states constituting
the branch S$^{\rm R1}_8$ \cite{CGPZ}. These short-scale dynamos are symmetric
about the vertical axis and stable to short-scale perturbations. We have shown
that, in a collective action, they are capable of generating a large-scale
magnetic field, whose geometry is determined by the properties
of the short-scale operator of linearisation around the respective steady state.
The operation of this large-scale dynamo is based on the joint action
of the combined eddy diffusivity and eddy advection, but it does not involve
the $\alpha$-effect. To the best of our knowledge, this mechanism is new.

Several remarks are in order.

1$^\circ$. We have found that only two small-scale neutral stability modes
of a steady convective dynamo with such a group of symmetries have
non-vanishing means,
each involving a combination of a mean horizontal velocity and magnetic field.
The solenoidality in slow variables of the two mean fields implies that only
a specific linear combination of these two modes is involved in the asymptotic
expansion of a large-scale perturbation, depending on a slow spatial
variable and evolving in the slow time. For a given state $\bf W$
from the considered branch, two such linear combinations can be determined.
Two other neutral small-scale stability modes, $\partial{\bf W}/\partial x_i$,
existing due to translation invariance in horizontal directions of equations
of convective dynamo, are zero-mean, i.e., their amplitudes
are not associated with any mean fields. The two amplitudes are essential
for the large-scale dynamics of the perturbation: if they are both set to zero,
then, by virtue of equations \rf{xi1}, the amplitude of the remaining
small-scale neutral mode, involving non-zero mean-field components,
also vanishes.

This clearly shows that omitting amplitudes of neutral zero-mean small-scale
stability modes (when they exist) in the description of the dynamics
of multiscale perturbations, which is the common practice in the theory
of mean-field electrodynamics, can render
the description inadequate. Such modes are not rare; they are featured
by translation-invariant and/or time-invariant MHD systems, and can be obtained
by differentiating the states, whose stability is analysed,
in the spatial variable along the invariant direction and/or time.

2$^\circ$. Requirements for the averaging procedures that can be applied
also differ significantly between MFE and MST. In MFE, any averaging is
deemed acceptable when analysing MHD turbulence, provided it satisfies
the Reynolds rules. Accordingly, averaging over a horizontal periodicity cell
(or, equivalently, over entire horizontal planes) is often considered
in literature. Perhaps, such averaging could be justified in the framework
of statistical approach to turbulence (e.g., for ensembles of turbulent eddies).
By contrast, our derivation of amplitude equations, following the MST approach
and using the PDE homogenisation techniques, shows that the only appropriate
averaging when considering two-scale perturbations of laminar MHD regimes
is over a periodicity domain of the perturbed state; for time-periodic perturbed
states, this must be supplemented by averaging over the temporal period.

Thus, a mathematically rigorous asymptotic procedure leaves no freedom
for choosing averaging to our taste; using spatial (or spatio-temporal)
averaging is inevitable --- this follows from the requirement of solvability
of equations in fast variables. Consequently, the MFE approach to such problems
(e.g., kinematic dynamo problems concerning generation of large-scale magnetic
field by a periodic array of small-scale flow cells) may yield quantitatively incorrect
results, when other types of averaging are used, even when they satisfy
the Reynolds rules and are allowed in MFE. Furthermore, no conclusions
of potential interest for astrophysics can be drawn from the results
of \cite{CH6,CH8} on the significant (i.e., responsible for the evolution
of large-scale magnetic field) $\alpha$-effect, because averaging over half
a periodicity cell performed in these works does not satisfy the Reynolds rules
(the significant $\alpha$-effect {\it ibid.} is zero).

3$^\circ$. The system of amplitude equations that we have derived and studied
numerically involves terms describing eddy diffusivity and eddy advection,
but not the $\alpha$-effect. This calls for a comparison of the action
of these mechanisms of magnetic field generation important in astrophysics.

The spectrum of the $\alpha$-effect operator
${\bf H}\mapsto\nabla\times{\mathfrak A}\bf H$, acting, for instance,
on space-periodic fields $\bf H$, is symmetric about the imaginary axis
\cite{Vi} (see also \cite{vz}, section~3.3.2). Under the action
of this operator, a generic perturbation experiences
a superexponential growth in slow time, which in this case is $T=\varepsilon t$
(while in the absence of the $\alpha$-effect large-scale amplitudes
evolve in O($\varepsilon$) slower time $T=\varepsilon^2t$, see
section~\ref{ord2}). In the MST framework,
when ${\mathfrak A}\ne0$, the $\alpha$-effect operator typically
controls the evolution of mean fields of perturbation solely (see \cite{vz},
sections~3.3.2, 4.2.3, 6.3.2, 7.3.2, 8.3.2). The superexponential growth might
be restrained by diffusion, since it is governed by a higher-order differential
operator. However, the orders of the $\alpha$-effect and eddy diffusivity
differential operators being different, the operators arise in solvability
conditions at different orders of the scale ratio and do not coexist.
In principle, emergence of the operators of $\alpha$-effect and eddy
diffusivity together is not ruled out in the MFE theory, where the so-called
$\beta$-effect is sometimes considered, in whose presence the mean e.m.f.~becomes
$$\la{\bf V}\times{\bf H}\ra=\sum_{i,j}\alpha_{ij}H_j{\bf e}_i
+\sum_{i,j,k}\beta_{ijk}{\partial H_j\over\partial X_i}{\bf e}_k.$$
However, this equality is only partially justified
in the MFE theory by considering the truncated Taylor expansion of the mean
field in the integral operator relating the fluctuating part of the magnetic
field to the mean part, and assuming that the respective kernels decay fast
(see \cite{Ra}). Whether this justification can be extended to a formal
asymptotic argument is an interesting open question.

Thus, to the best of our knowledge there are no mathematically solid evidences
that in multiscale problems, where a rigorous derivation of amplitude equations
is possible, the action of the $\alpha$-effect operator can be restrained
by eddy diffusivity or the mechanisms of nonlinear saturation.
Hence, it can be described as essentially
``self-destructing'': In a system with the $\alpha$-effect, the perturbation
grows superexponentially until it loses the asymptotic smallness and thereby
significantly affects the perturbed state. In other words, the $\alpha$-effect
can generate magnetic field for a relatively short time, but
it destabilises the existing small-scale structure
and modifies the entire MHD system into a completely new regime. For this new
regime, the analysis of eddy effects must be done anew (provided the scale
separation is preserved and the multiscale analysis can be implemented;
actually, it is also not guaranteed that the new regime will be capable
of generating the field). This is incompatible with the astrophysical reality
--- large-scale cosmic magnetic fields are known to exist for very long times.
Therefore, it is natural to expect that a sequence of such events
of self-destruction terminates at an MHD state, where the magnitude
of the $\alpha$-effect is sufficiently small not to cause further disturbances.

Within this line of reasoning, the feasibility of the $\alpha$-effect
paradoxically, may owe to the phenomenon, originally perceived as significantly
reducing the importance of the $\alpha$-effect, namely, the $\alpha$-quenching,
i.e., a drastic decrease of the $\alpha$-effect on increasing the magnetic
Reynolds number,
discovered by S.I.~Vainshtein \cite{VC}. In the kinematic dynamo problem
for a flow with an external scaling, the $\alpha$-effect operator is present
in the equation for mean fields together with molecular diffusion \cite{Vi}
(see also \cite{vz}, ch.~10, 11). In astrophysics, molecular diffusion is weak.
However, if the $\alpha$-quenching inhibits the action of the $\alpha$-effect
to the levels, where it is essentially offset by molecular magnetic diffusivity,
then, at least within this model (also mathematically rigorous),
the $\alpha$-effect operator may contribute to stationary magnetic field
generation without causing the restructuring of the MHD system hosting it.

As our results show (see section~\ref{resu}), in the absence
of the $\alpha$-effect the overall character of the large-scale evolution
is not that different. If negative combined eddy diffusivity is present,
then a generic perturbation grows superexponentially. As mentioned above,
the characteristic slow time
of the large-scale evolution is then slower than for the $\alpha$-effect, but
the ultimate result of the evolution is the same: destruction of the underlying
MHD regime and the onset of a new one (which can lack scale separation).
We have no evidence that taking into account nonlinear terms
in equations for weakly nonlinear perturbations can halt such processes. Here
we have studied numerically the case, where the action of the combined eddy
diffusivity is stabilising (as that of molecular diffusivity). In this case
the large-scale evolution significantly depends on how large is the initial
perturbation. When it is below a certain threshold, the amplitudes tend
to constants and the perturbation evolves to a linear combination
of the two zero-mean neutral modes; otherwise, it blows up in apparently
a finite time. Therefore, like in the presence of the $\alpha$-effect,
the large-scale evolution controlled by eddy diffusivity and eddy advection
can perturb the original small-scale dynamics till its structure is
significantly modified, but the growth rate of these processes is much,
O($\varepsilon$), slower.

An open question is whether the system of amplitude equations
\rf{xi1}--\rf{xi3} and \rf{xi4} has a solution which is a travelling wave.
We have failed to find this regime for the steady dynamo
at ${\rm Ta}=675$, but this can be due to the scarceness of our efforts; also,
it can, in principle, exist for perturbations of other steady states from
the branch that we have considered, or in other branches with the symmetries
compatible with our multiscale analysis (such as the time-periodic dynamo
P$^{\rm R1}_4$). If such a solution exists and is stable, it would be
an example of a large-scale perturbation that does not act self-destructively
as just discussed; if it is unstable in slow variables, it is nevertheless
of interest as an invariant object in the phase space influencing the dynamics
of amplitudes.

4$^\circ$. We have analysed numerically the system of amplitude equations
derived here for convective hydromagnetic regimes with the symmetry group of
states in the branch
S$^{\rm R1}_8$. The behaviour of its solutions is complex, because this is
a mixed system, involving both evolutionary and non-evolutionary nonlinear
equations. Many questions remain open, for instance: Does this system allow
patterns of behaviour of perturbation, different from those that we have
identified? Does the blow-up occur for other boundary conditions?
Does the solution for ${\rm Ta}=675$ ``converge'' at the blow-up time
to a distribution involving the $\delta$-function, as computations suggest?
If this $\delta$-function is interpreted as a derivative of certain quantities
experiencing a jump, can these quantities be identified in physical terms?
How can a solution be continued in some weak sense beyond the blow-up time
$T_\star$?

5$^\circ$. The system of amplitude equations that we have derived is very
restrictive: it can only be used to describe weakly nonlinear perturbations,
where amplitudes are initially constant on lines, perpendicular to certain
well-defined
directions $\bf q$ on the plane of slow variables. This system, therefore,
does not describe general two-scale perturbations of a convective MHD state,
that are of the form \rf{vs}--\rf{ts} and obey equations \rf{pNS}--\rf{pso}.
An interesting open question is to construct an asymptotic formalism
for perturbations of a more general form.

6$^\circ$. Finally, it is of interest to perform similar investigations
for other branches of short-scale convective hydromagnetic regimes:
(i) for time-periodic MHD states constituting branch P$^{\rm R1}_4$ that
emerge in the interval $704<{\rm Ta}<705$ in a Hopf bifurcation
from S$^{\rm R1}_8$ considered here (large-scale perturbations
of time-periodic states involve an additional amplitude);
(ii) for steady states and time-periodic regimes constituting the branches,
found in \cite{CGPZ}, that possess other groups of symmetries, for which $\U=0$.

\section*{Acknowledgments}

We are grateful to Erico Rempel for discussions.
RC acknowledges financial support from FAPESP (Brazil, grant 2013/01242-8).

\end{document}